\documentclass[
 reprint,prl,
 superscriptaddress,
 amsmath,amssymb,
 aps
]{revtex4-2}

\usepackage[utf8]{inputenc}
\usepackage{mathtools}
\usepackage{amsmath,amssymb,mathrsfs}
\usepackage{xcolor}
\usepackage{graphicx}
\definecolor{mycol}{RGB}{0,102,204}
\usepackage[colorlinks, linkcolor=mycol, citecolor=mycol, urlcolor=black, breaklinks]{hyperref}
\usepackage[capitalise]{cleveref}

\newcommand{\dd}{{\rm d}}
\newcommand{\im}{{\rm Im\,}}
\newcommand{\re}{{\rm Re\,}}
\newcommand{\tr}{{\rm Tr}}

\newcommand{\E}{\mathbb{E}}

\begin{document}

\title{Exact solutions of non-Markovian dissipative Ising models}

\author{Mohammad Maghrebi}
\email{maghrebi@msu.edu}
\affiliation{Department of Physics and Astronomy, Michigan State University, East Lansing, Michigan 48824 USA}

\begin{abstract}
Applications of quantum devices and simulators require an accurate understanding of their dynamics as well as  the effects of environmental noise. While some sources of noise can be modeled as Markovian dissipation, coupling to resonators or vibrational modes often produces non-Markovian dynamics as well, making it difficult to characterize the system, particularly for interacting many-body systems. Here, we identify a broad class of exactly solvable many-body quantum systems with both non-Markovian dynamics and Markovian dissipation. We consider a prototypical model of spins coupled to lossy bosonic modes that mediate Ising-type interactions. We exclude a transverse field but allow local spin dissipation that need not commute with the interactions, including dephasing, relaxation, pumping, and even processes that generate coherences. Using an exact stochastic representation, we decouple the many-body dynamics into the evolution of individual spins, yielding exact spin observables and correlations without adiabatically eliminating the bosons. In the Markovian limit, our solutions describe Ising models with correlated dephasing and nonreciprocal interactions. An example of spin squeezing shows that both the achievable squeezing and its optimal preparation time depend strongly on the interplay between non-Markovian dynamics and Markovian dissipation.
\end{abstract}

\maketitle

Understanding the dynamics of open quantum systems is increasingly important for quantum devices aimed at preparing metrologically useful states or simulating many-body physics~\cite{bohnetQuantumSpinDynamics2016, altmanQuantumSimulatorsArchitectures2021a,monroeProgrammableQuantumSimulations2021}. In these devices, coupling to cavity or vibrational modes can introduce non-Markovian dynamics, while unavoidable environmental noise is often modeled as Markovian dissipation~\cite{kaerNonMarkovianModelPhotonAssisted2010,nagyNonequilibriumQuantumCriticality2015,cosacchiPathintegralApproachNonequilibrium2018,lundgrenNatureNonequilibriumPhase2020,lemmerTrappedionSimulatorSpinboson2018,soTrappedionQuantumSimulation2024,soQuantumSimulationCharge2025,kamarSpinbosonModelDephasing2024,sunQuantumSimulationSpinboson2025a}. Despite considerable progress in theoretical and computational approaches ~\cite{devegaDynamicsNonMarkovianOpen2017,xuColloquiumSimulatingNonMarkovian2026,keelingProcessTensorApproaches2026}, characterizing non-Markovian dynamics is particularly challenging in interacting many-body systems. This calls for accurate analytical results and, ideally, exact solutions to benchmark numerical approaches and characterize noise in quantum devices \cite{gandhariQuantumNonMarkovianNoise2026a}. Yet exactly solvable models with non-Markovian dynamics as well as Markovian dissipation remain  scarce ~\cite{betzholzQuantumOpticalMaster2014,johnssonModifiedCoherenceQuantum2022}, especially in many-body systems.
\begin{figure}[t!]
    \centering
    \includegraphics[width=0.95\columnwidth]{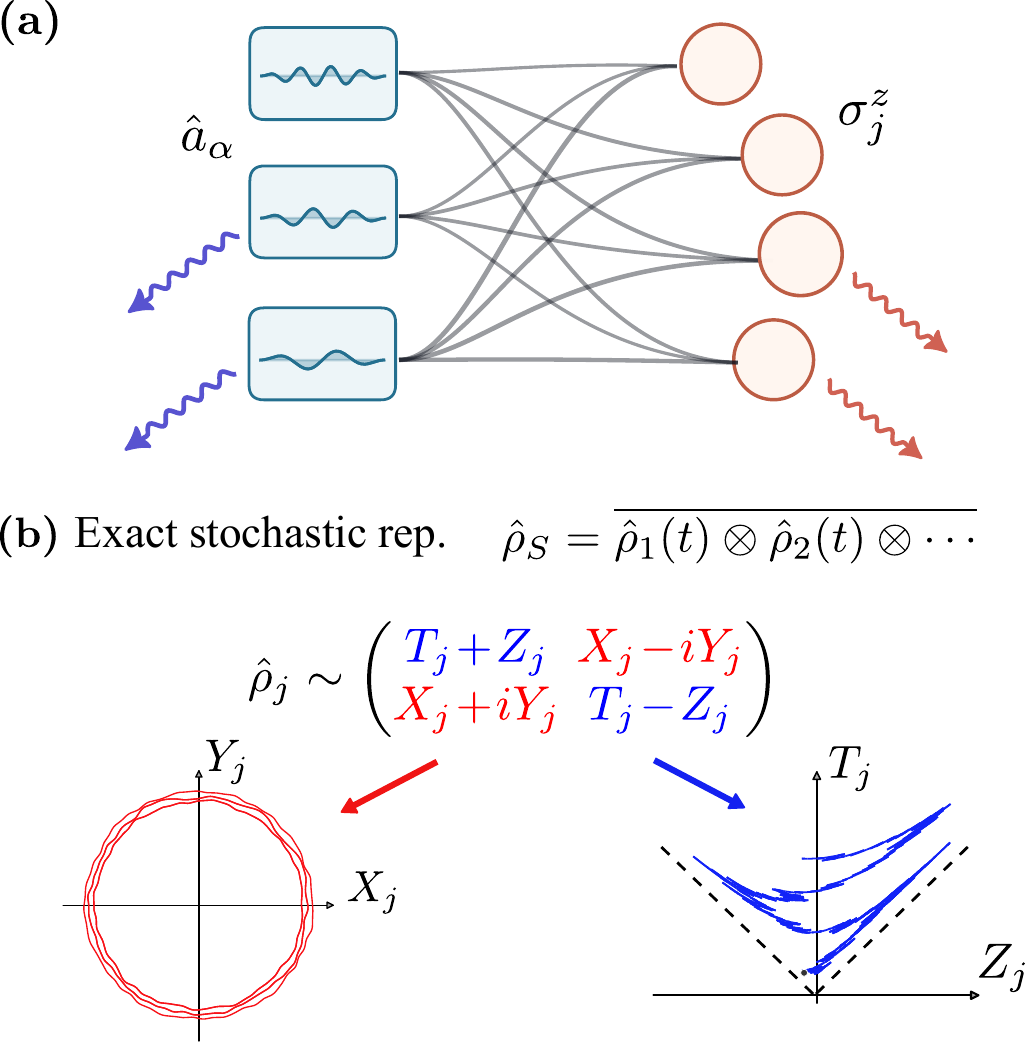}
    \caption{Schematic model and exact stochastic solution. (a) Spins coupled to lossy bosonic modes and subject to local spin dissipation.
    The bosons mediate Ising-type interactions and give rise to non-Markovian dynamics. (b)~The reduced density matrix of the spins is obtained by averaging the product of individual stochastic density matrices which are generally neither trace-1 nor positive. Without a transverse field or local spin dissipation, their matrix elements form two closed, exactly solvable pairs: their transverse components \((X_j,Y_j)\) undergo stochastic rotations, while the evolution of the diagonal components \((T_j,Z_j)\) resembles stochastic Lorentz boosts, as shown schematically. Local dephasing, relaxation and pumping and a class of coherence-generating processes can also be incorporated while retaining exact solvability. Noise averaging yields exact observables and correlations without adiabatic elimination.}
    \label{fig:1}
\end{figure}

In this work, we identify a broad and experimentally relevant class of exactly solvable many-body open quantum systems with non-Markovian dynamics.
We consider a prototypical model of spins coupled to lossy bosonic modes that mediate Ising-type interactions. We exclude a transverse field, but allow local spin dissipation that need not commute with the Ising interaction, including spontaneous emission, pumping, and even processes that generate coherences.
To obtain exact solutions, we utilize a stochastic approach recently developed in Ref.~\cite{kamarHybridQuantumClassicalStochastic2025a}, which decouples the many-body dynamics into the evolution of individual spins. This yields exact spin observables and correlations without adiabatically eliminating the bosons. We provide an example of  spin squeezing where non-Markovian dynamics combined with spin dissipation qualitatively alters the dynamics. In the Markovian limit, our exact solutions also describe dissipative Ising models with correlated dephasing and nonreciprocal interactions.

\textit{Model.---}We consider an open quantum system governed by the quantum master equation
\begin{equation}
\frac{\dd}{\dd t}\hat\rho
=-i[\hat H,\hat\rho]
+{\cal D}_B(\hat\rho)
+{\cal D}_S^{\rm loc}(\hat\rho),
\end{equation}
where \(N\) spins, indexed by \(j\), are coupled to lossy bosonic modes, indexed by \(\alpha\):
\begin{align}
&\hat H=
\sum_\alpha\omega_\alpha\hat a_\alpha^\dagger\hat a_\alpha
+\frac{1}{2\sqrt N}\sum_{\alpha j}\sigma_j^z
\left(g_{\alpha j}\hat a_\alpha
+g_{\alpha j}^*\hat a_\alpha^\dagger\right),\nonumber\\
&{\cal D}_B(\hat\rho)=
\sum_\alpha\kappa_\alpha
\left(2\hat a_\alpha\hat\rho\hat a_\alpha^\dagger
-\{\hat a_\alpha^\dagger\hat a_\alpha,\hat\rho\}\right).
\end{align}
The Hamiltonian describes Rabi-type couplings with generally complex and nonuniform strengths \(g_{\alpha j}\), while \({\cal D}_B\) describes bosonic loss. Additionally, the spins are subject to local dissipation,
\(
{\cal D}_S^{\rm loc}(\hat\rho)
=\sum_{j,\mu}\Gamma_{\mu}{\cal D}[\hat L_{j,\mu}]\hat\rho,
\)
where
\(
{\cal D}[\hat L]\hat\rho
=\hat L\hat\rho\hat L^\dagger
-\frac12\{\hat L^\dagger\hat L,\hat\rho\}
\), and \(\mu\) labels different dissipative processes. See \cref{fig:1}(a) for a schematic representation of the model.

For a single bosonic mode and uniform coupling, we recover the familiar open Dicke model \cite{dimerProposedRealizationDickemodel2007} but without a transverse field. Although the Hamiltonian is diagonal in the spin \(\sigma^z\) basis, the nonequilibrium dynamics remains nontrivial, especially in the presence of non-commuting dissipation. In the regime where adiabatic elimination is valid, we obtain an effective spin model with correlated dephasing and, for complex \(g_{\alpha j}\), nonreciprocal Ising interactions~\cite{metelmannNonreciprocalPhotonTransmission2015a,seifDistinguishingQuantumClassical2022}
as discussed below.

\textit{Exact stochastic approach.---}To obtain the exact solutions, we use a stochastic approach that decouples spins from one another and from the bosons, while keeping the dynamics local in time. To illustrate this approach, we consider a single bosonic mode and drop the index $\alpha$, writing $\omega_\alpha, \kappa_\alpha, g_{\alpha j}\to \omega,\kappa, g_j$. The generalization to multiple modes is  straightforward.
Following Ref.~\cite{kamarHybridQuantumClassicalStochastic2025a}, the bosonic operator $\hat a$ is traded for a $c$-number stochastic field $\psi(t)$, satisfying the semiclassical equation for the cavity mode alone, with no explicit coupling to spins,
\begin{subequations}\label{eq:stoch_1}
    \begin{align}
        i\dot\psi(t) =  (\omega-i\kappa)\psi(t)
    +\frac{1}{\sqrt{N}}\sum_{j=1}^N\xi_j(t)\,.
    \end{align}
Here, $\xi_j$ are independent complex white noise variables, 
\(
    \overline{\xi_i(t)\xi_j^*(t')}=\kappa\delta_{ij}\delta(t-t'),
\)
 $\overline{\xi_i(t)\xi_j(t')}=0$.
Introducing independent noise variables for different spins allows us to \textit{stochastically} decouple their dynamics. Each spin then evolves as
    \begin{align}
    \begin{split}\label{eq:2b}
    \frac{\dd }{\dd t} \hat \rho_j =& {\cal L}_j(\hat \rho_j) -  \frac{i}{2\sqrt{N}}(g_j\psi +g_j^*\psi^*) [\sigma_j^z,\hat\rho_j]  \\
    &\quad \quad\,\,\,\, + \frac{1}{4\kappa} (g_j \xi_j +g_j^* \xi_j^{*}) \{\sigma^z_j, \hat \rho_j\}\,,
    \end{split}
    \end{align}
\end{subequations}
where $\hat\rho_j$ is the stochastic density matrix describing spin $j$. Here,  ${\cal L}_j$ captures the single-spin Liouvillian, which can in principle include a transverse field, although we exclude it to obtain the exact solutions below. For a single spin, \cref{eq:2b}  resembles the stochastic Liouville–von Neumann equation \cite{stockburgerExactNumberRepresentation2002,xuColloquiumSimulatingNonMarkovian2026}. However, the latter assumes a Gaussian initial bosonic state and generally does not preserve hermiticity for each trajectory; see \cite{kamarHybridQuantumClassicalStochastic2025a} for further discussion.

To compute physical quantities of interest, we average over the stochastic solutions. In particular, the reduced density matrix for spins, $\hat\rho_S=\tr_B(\hat\rho)$, is given by the stochastic average of the product of individual density matrices,
\begin{equation}\label{eq:average}
    \hat \rho_S(t) = \overline{\otimes_j \hat \rho_j(t)}\,,
\end{equation}
where the average is over white noise as well as the initial Wigner function of the bosonic mode.
Expectation values of spin observables can then be computed as $\langle \hat O_S\rangle=\tr_S(\hat \rho_S \hat O_S)$. We emphasize that this approach is exact, with no approximations involved; we provide a simple proof in  the Supplemental Material (SM) \cite{sm}. Explicitly constructing \(\hat\rho_S\), however, requires an exponentially large matrix. For typical observables that are products over spins, $\hat O_S = \otimes_j \hat O_j$, we can instead compute the single-spin traces and then average their product, $\langle \hat O_S\rangle =\overline{\prod_j\tr(\hat \rho_j \hat O_j)}$.

At first glance, the product form in \cref{eq:average} might suggest that the state is separable and cannot capture entanglement. However, the density matrices $\hat \rho_j$s need not be physical for individual trajectories
as the term proportional to the noise involves an anticommutator.
One can see that $\hat\rho_j$s, while hermitian, are not generally trace-1 or positive.
Without a constraint on positivity, \cref{eq:average} can describe a general, possibly entangled, state.
Furthermore, since the density matrices are not trace-1, their traces should be retained when computing the stochastic averages. For example, the expectation value of an observable $\hat O_j$ for spin $j$ involves the traces of the other density matrices:
\begin{equation}\label{<Oj>}
    \langle \hat O_j\rangle = \overline{\tr(\hat O_j \hat \rho_j)\prod_{l\ne j}\tr(\hat\rho_l)}\,.
\end{equation}

\textit{Exact solutions from Bloch equations.---}Since the spins are effectively decoupled in our stochastic approach, it is convenient to write each spin's density matrix in the Bloch sphere representation,
\begin{equation}\label{eq:Bloch_0}
    \hat \rho_j = \frac{T_j \boldsymbol{1}_j+ X_j\sigma^x_j+ Y_j \sigma^y_j+ Z_j\sigma^z_j}{2},
\end{equation}
where $\boldsymbol{1}_j$ is the corresponding identity matrix with a coefficient $T_j=\tr(\hat \rho_j)$ which we allow to be different from 1 since the stochastic dynamics is not trace preserving. Note that the stochastic dynamics preserves hermiticity, hence real-valued (stochastic) variables $T_j,X_j,Y_j,Z_j$.

For simplicity,
we first assume real couplings \(g_j\) and set \(\mathcal L_j=0\), neglecting spin dissipation.
Plugging \cref{eq:Bloch_0} into \cref{eq:2b}, we find the stochastic Bloch-like equations,
\begin{equation}\label{eq:Bloch}
\begin{split}
    \dd X_j &= - h_j(t) Y_j \dd t, \qquad
    \dd Y_j =  h_j(t) X_j  \dd t, \\
    \dd Z_j &= \frac{g_j}{\sqrt{2\kappa}} T_j \dd W_j,\qquad
    \dd T_j =  \frac{g_j}{\sqrt{2\kappa}} Z_j \dd W_j,
\end{split}
\end{equation}
where we have defined $h_j(t)\equiv  \frac{g_j}{\sqrt{N}}(\psi+\psi^*)$.
We write these equations in the Itô convention using the Wiener increments
satisfying \(\dd W_i\,\dd W_j=\dd W_i'\,\dd W_j'=\delta_{ij}\,\dd t\) and \(\dd W_i\,\dd W_j'=0\).
Paralleling \cref{<Oj>}, the average of an observable, say $\sigma^a_j$, takes the form
\begin{equation}\label{eq:E}
    \langle \sigma^a_j(t)\rangle = \E [X^a_j(t) \prod_{l\ne j} T_l(t)]\,,
\end{equation}
where $\E$ denotes the average of the random process, and $X^a= X,Y,Z$ for $a=x,y,z$.

For each spin, the equations for $X_j, Y_j$ form a closed pair, as do those for $Z_j, T_j$. The first line in \cref{eq:Bloch} describes rotations in the $X$-$Y$ plane under a longitudinal field $h_j(t)$, while the evolution of $Z_j,T_j$ is governed by diffusion terms (proportional to noise), and resembles a (stochastic) Lorentz boost; see \cref{fig:1}(b). \Cref{eq:Bloch} can then be solved exactly as
\begin{equation}\label{eq:solution}
 \begin{split}
        &\!\!\!\!X_j(t)+iY_j(t) = e^{i\int_0^t dt' h_j(t')} (X_j(0)+iY_j(0))\,, \\
        &\!\!\!\! T_j(t) \pm Z_j(t)=
e^{-\frac{|c_j|^2}{2}t \pm c_j W_{j}(t)}
(T_j(0) \pm Z_j(0))\,,
\end{split}
\end{equation}
with $c_j \equiv \frac{g_j}{\sqrt{2\kappa}}$, and $W_j(t)=\int_0^t \dd W_j(s)$ a Wiener process (standard Brownian motion). The linear stochastic equation for $\psi$ is also solved in terms of noise using the causal Green's function of the bosonic mode, $G(t)=-i\Theta(t)  e^{-i\omega t-\kappa t}$.
Generalization to complex $g_j$ follows immediately \cite{sm}.

The above equation provides an exact solution to the full dynamics, and can be used to compute spin observables of interest. As an example, we consider spins initially polarized along the positive $x$ direction and the boson described by the Wigner function ${\mathscr W}_0(\psi_0)$. We can then evaluate the Gaussian averages over noise in \cref{eq:E} to obtain
\label{eq:full}
    \begin{align}\label{eq:X+iY}
         \langle  \sigma^x_j +i \sigma^y_j \rangle_t  = \mathcal A_{j}(t)e^{-\Lambda_{jj}(t) } \prod_{l\ne j} \cos\Theta_{jl}(t)\,,
    \end{align}
where we have defined
\begin{subequations}
\begin{align}
    \mathcal A_j(t)&=\int \dd^2\psi_0 {\mathscr W}_0(\psi_0) e^{i \frac{1}{\sqrt{N}} (i{g_j\mathscr G}(t)\psi_0+{\rm c.c.})}\,,\\
    \Lambda_{jj}(t)&=\frac{\kappa}{N}|g_j|^2 \int_0^t \dd s |{\mathscr G}(s)|^2\,, \\
    \Theta_{jl}(t)&=\frac{1}{N} \int_0^t \dd s \re( g_j g_l^*{\mathscr G}(s))\,,
\end{align}
\end{subequations}
with ${\mathscr G}(t) \equiv \int_0^t \dd s G(s)$ and generally complex couplings $g_j$.
Here, \(\mathcal A_j(t)\) captures the contribution from the initial bosonic state,  \(\Lambda_{jj}(t)\) describes an effective dephasing due to cavity loss, and \(\Theta_{jl}(t)\) encodes the boson-mediated interactions between spins. Our exact solutions capture the full non-Markovian dynamics through \(\mathscr G(t)\), without requiring adiabatic elimination.
Including multiple bosonic modes is straightforward with the substitution $g_{j}\to g_{\alpha j}, \mbox{and } G,{\mathscr G}\to G_\alpha, {\mathscr G}_\alpha$ and summing the expressions for $\Lambda_{jj}$ and $\Theta_{jl}$ over $\alpha$, while integrating over the joint Wigner function in ${\cal A}_j$.
Correlation functions involving multiple spins as well as bosons can be computed similarly; details of the above calculations are provided in the SM \cite{sm}.

\textit{Exact solution with spin dissipation.---}We now include local spin dephasing, relaxation, and pumping all in the $z$ basis via the jump operators \(\hat L=\sigma^z,\sigma^\mp\) and rates \(\Gamma_\phi,\Gamma_\downarrow,\Gamma_\uparrow\), respectively; the raising and lowering spin operators are defined as $\sigma^\mp\equiv (\sigma^x\mp i\sigma^y)/2$.
The stochastic equations still form two closed pairs: The transverse components \(X_j,Y_j\) in \cref{eq:solution} just acquire an overall exponential decay, \(\exp(-\Gamma't)\), with \(\Gamma'=2\Gamma_\phi+(\Gamma_\downarrow+\Gamma_\uparrow)/2\). The stochastic differential equation for $T_j$ remains the same, while that for \(Z_j\) acquires the local contribution
\(
\dot Z_j|_{\rm loc}
=-(\Gamma_\downarrow-\Gamma_\uparrow)T_j
-(\Gamma_\downarrow+\Gamma_\uparrow)Z_j.
\)
But a closed form solution for $T_j,Z_j$ is no longer generally available.

Nevertheless, the stochastic averages of interest can be obtained from simple ordinary differential equations.
To this end, we first note that \cref{eq:X+iY} retains its form,
\begin{equation}\label{eq:10}
    \left\langle\sigma_j^x+i\sigma_j^y\right\rangle_t
    =
    \mathcal A_j(t)e^{-\Gamma't-\Lambda_{jj}(t)}
    \prod_{l\ne j}p_{jl}(t),
\end{equation}
with  $\mathcal A_j$ and $\Lambda_{jj}$ unchanged, and a product over the newly introduced factors $p_{jl}$. Each factor $p_{jl}$ can be interpreted as the average of the phase factor accumulated by spin \(j\) due to the spin-flip history of spin $l$. More precisely,
\(
    p_{jl}(t)=\mathbb E_{\rm flips}[e^{i\phi_{jl}(t)}]
\), where the phase $\phi_{jl}$ is defined as
\begin{equation}\label{eq:k_jl}
    \phi_{jl}(t)=\int_0^t \dd s\,k_{jl}(t-s)z_l(s),
\end{equation}
with $z_l(s)=\pm1$ independent classical spins that flip down
at rate $\Gamma_\downarrow$ and up at rate $\Gamma_\uparrow$, and   $k_{jl}(t)\equiv
\dot \Theta_{jl}(t)=\re[g_jg_l^*\mathscr G(t)]/N$.
To solve for $p_{jl}(t)$, let $p_{jl}^{\uparrow/\downarrow}(t)$ denote the phase averages conditioned on the spin $l$ being initially up or down respectively. As we show in the SM~\cite{sm}, these factors satisfy the equation
\begin{equation}\label{eq:14}
    \frac{\dd}{\dd t}
    \begin{pmatrix}
        p_{jl}^\uparrow\\
        p_{jl}^\downarrow
    \end{pmatrix}
    =
    \begin{pmatrix}
        ik_{jl}(t)-\Gamma_\downarrow & \Gamma_\downarrow\\
        \Gamma_\uparrow & -ik_{jl}(t)-\Gamma_\uparrow
    \end{pmatrix}
    \begin{pmatrix}
        p_{jl}^\uparrow\\
        p_{jl}^\downarrow
    \end{pmatrix},
\end{equation}
with $p_{jl}^{\uparrow/\downarrow}(0)=1$. The initial polarization along $+x$ gives equal weights to the two orientations, so $p_{jl}=(p_{jl}^\uparrow+p_{jl}^\downarrow)/2$.
The coherence of spin \(j\) thus probes the other spins’ telegraph fluctuations through the accumulated phase, as in the Markovian case considered by McDonald and Clerk~\cite{mcdonaldExactSolutionsInteracting2022}. Here, however, the kernel \(k_{jl}\) in \cref{eq:k_jl} also captures non-Markovian memory effects.
While a closed form solution to \cref{eq:14} is generally not available, it can be easily solved numerically.
Two- (or higher order) spin correlators can be obtained similarly \cite{sm}, and will be used to calculate squeezing below.

\emph{Coherence-generating dissipation.---}We can also incorporate dissipation that generates coherence while preserving exact solvability. An example is resetting each spin to \(|+x\rangle\), described by the master equation \cite{hartmannSteadystateEntanglementOpen2006}
\[
\frac{\dd }{\dd t} {\hat\rho}|_{\rm reset} =r\sum_j \big[|+x\rangle_j\langle+x|\otimes \tr_j\hat \rho-\hat \rho\big]\,.
\]
Under this dynamics, the populations in the \(z\) basis evolve independently of the coherences~\cite{foss-feigSolvableFamilyDrivenDissipative2017}, so the stochastic equations for \(T_j,Z_j\) remain closed. The reset term damps \(X_j,Y_j\) equally, while the diagonal components provide a source for the coherence. Defining \(C_j=X_j+iY_j\), its stochastic evolution is given by
\begin{equation}
\dd C_j=[ih_j(t)-r]C_j\,\dd t+r T_j\,\dd t.
\end{equation}
The required stochastic averages again reduce to simple ordinary differential equations,
which capture the non-Markovian dynamics, thus extending our exact solutions to coherence generating dissipation; see the SM \cite{sm} for further details.

\textit{Nonreciprocal Ising model.---}At large cavity loss or frequency, bosons can be adiabatically eliminated resulting in a spin-only model described by an Ising Hamiltonian and correlated dephasing \cite{zouSpatiallyCorrelatedClassical2024b,jachinowskiSpinonlyDynamicsMultispecies2026},
\begin{align}\label{eq:Ising}
\begin{split}
    &\hat H_S = \sum_{j<l} J_{jl}\sigma^z_j \sigma^z_l,  \\
    &{\cal D}_S^{\rm corr}(\hat \rho_S)=
    \sum_{j,l} \Gamma_{jl}\big(\sigma^z_j \hat \rho_S \sigma^z_l- \frac{1}{2}\{\sigma^z_j\sigma^z_l,\hat \rho_S\}\big)\,,
\end{split}
\end{align}
with the subscript $S$ denoting spin-only (super)operators; the full dissipator contains both local and correlated dissipation, ${\cal D}_S={\cal D}^{\rm loc}_S+ {\cal D}^{\rm corr}_S$. The effective couplings are given by \cite{sm}
\begin{align}
\begin{split}
    J_{jl} &= -\frac{1}{2N}\sum_\alpha \frac{\omega_\alpha}{\omega_\alpha^2+ \kappa_\alpha^2} \re(g_{\alpha j} g_{\alpha l}^*), \\
    \Gamma_{jl}&=\frac{1}{2N}\sum_\alpha \frac{\kappa_\alpha}{\omega_\alpha^2+ \kappa_\alpha^2} g_{\alpha j}^* g_{\alpha l}\,.
\end{split}
\end{align}
Note that the matrix $\Gamma_{jl}$ is manifestly hermitian and positive semidefinite, but for complex $g_{\alpha j}$, it is generally not real, resulting in nonreciprocal dynamics~\cite{seifDistinguishingQuantumClassical2022}.

Exact solutions for the resulting dissipative Ising model immediately follow by taking the proper limit. Specifically, in this limit we find
\[
k_{jl}(t)\longrightarrow 2J_{jl}^{\rm eff},
\]
in which case \cref{eq:14} can be solved in closed form.
Here, \(J_{jl}^{\mathrm{eff}}=J_{jl}-\im\Gamma_{jl}\) defines the action of spin \(l\) on spin \(j\) that is generally different from that of $j$ on $l$, \(J_{lj}^{\mathrm{eff}}=J_{jl}+\im\Gamma_{jl}\), giving rise to nonreciprocal dynamics.
For the boson initially in its vacuum state and purely local dissipation ($\Gamma_{jl}=0$ for $j\ne l$), we recover the results in Refs.~\cite{foss-feigNonequilibriumDynamicsArbitraryrange2013a,foss-feigDynamicalQuantumCorrelations2013,mcdonaldExactSolutionsInteracting2022}. Furthermore, our approach extends those results to correlated dissipation ($\Gamma_{jl}\ne 0$) and nonreciprocal coupling ($\im\Gamma_{jl}\ne 0$).
Interestingly, we find that the reduced dynamics of an arbitrary subsystem \(\mathcal C\) depends on its couplings to the spins in its complement only through \(J_{jl}^{\rm eff}\), with \(j\in\mathcal C\) and \(l\notin\mathcal C\). The subsystem dynamics can then be obtained using reciprocal Ising couplings of strength \(J_{jl}^{\rm eff}\) across the boundary and no cross-boundary correlated dephasing, while keeping the internal dynamics of \(\mathcal C\) unchanged \cite{sm}.

\begin{figure}[t!]
    \centering
    \includegraphics[scale=0.5]{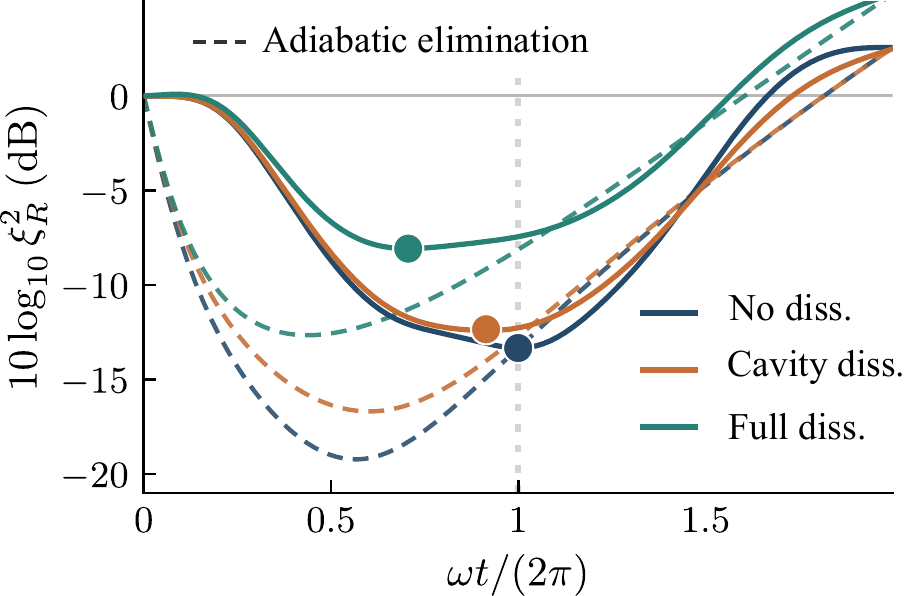}
    \caption{Spin squeezing with non-Markovian dynamics and Markovian dissipation. Exact (solid) and adiabatically eliminated (dashed) dynamics of the Wineland parameter are shown for \(N=1000\) spins uniformly coupled to a single bosonic mode with \(g/\omega=1.8\). Different curves represent no dissipation (blue), cavity loss only (orange), and cavity loss together with local spin dephasing, emission, and pumping (green). Local spin dissipation shifts the optimal squeezing (marked by solid circles) earlier and reduces the achievable squeezing. Spins initially point along \(+x\) and the bosonic mode is in its vacuum state.  Dissipation parameters are \(\kappa/\omega=0.05\) and  \(\Gamma_\phi/\omega=0.01\), \(\Gamma_\downarrow=\Gamma_\uparrow=\Gamma_\phi/2\), comparable to those reported in trapped-ion experiments \cite{bohnetQuantumSpinDynamics2016}.}
    \label{fig:2}
\end{figure}

\emph{Spin squeezing.---}The spin–boson coupling considered here is directly relevant to generating metrologically useful spin-squeezed states in AMO platforms such as trapped ions~\cite{bohnetQuantumSpinDynamics2016}. Spin squeezing beyond adiabatic elimination has been characterized exactly for unitary spin–boson dynamics~\cite{dylewskyNonperturbativeCalculationPhonon2016}. Our exact solutions  further allow us to include the experimentally relevant Markovian dissipation of both bosons and spins.
We consider a single mode with uniform real couplings, \(g_j=g\), initially in the vacuum, and spins polarized along \(+x\). The boson-mediated interaction generates spin squeezing, which we quantify by the Wineland parameter
\(
\xi_R^2
={N\min_\vartheta\operatorname{Var}(\hat S_y\cos\vartheta+\hat S_z\sin\vartheta)} /{\langle \hat S_x\rangle^2}
\)
with \(\hat S_\mu=\frac12\sum_j\sigma_j^\mu\) the total spin;
$\xi_R^2<1$ indicates metrologically useful squeezing \cite{winelandSqueezedAtomicStates1994}. \Cref{fig:2} shows the exact dynamics with and without cavity and spin dissipation and compares them against the prediction from adiabatic elimination. For the parameters shown, optimal squeezing without dissipation occurs at \(t=2\pi/\omega\) when the boson returns to its initial state and disentangles from spins \cite{dylewskyNonperturbativeCalculationPhonon2016}. Dissipation alters the coherent dynamics and substantially shifts the optimal time earlier, even though the dissipation rates are much smaller than \(\omega\).
In all cases, the dynamics from adiabatic elimination significantly deviates from the exact dynamics.
Accurately predicting optimal squeezing in this regime therefore requires accounting for both non-Markovian dynamics and Markovian dissipation.

\textit{Conclusion and Outlook.---}In this work, we have uncovered exact solutions for the non-Markovian dynamics of spins coupled to lossy bosons and subject to local spin dissipation.
An experimentally relevant example of spin squeezing illustrates how non-Markovian dynamics together with dissipation can have a nontrivial effect.
Here, we have begun to identify coherence generating spin dissipation that remains exactly solvable even beyond finite-range models \cite{foss-feigSolvableFamilyDrivenDissipative2017}, a topic that remains an interesting direction for future research.
Furthermore, our approach can be readily generalized to other models, such as interacting bosons whose population dynamics remains closed \cite{mcdonaldExactSolutionsInteracting2022,torresOptomechanicalDampingBasis2019}. Finally, even when analytical solutions are unavailable, our stochastic approach remains exact, and can provide a promising route to numerical simulations of more general non-Markovian many-body systems in the presence of Markovian dissipation \cite{kamarHybridQuantumClassicalStochastic2025a}.

\begin{acknowledgments}
The author acknowledges support from the National Science Foundation under NSF CAREER Award No.~DMR-2142866 and the Office of Naval Research under Grant No.~N000142612232. This project used GPT-6 Astra for  general checking, proofreading and assistance with the numerical calculations.
\end{acknowledgments}

\bibliography{exact_2}

\begin{thebibliography}{38}%
\makeatletter
\providecommand \@ifxundefined [1]{%
 \@ifx{#1\undefined}
}%
\providecommand \@ifnum [1]{%
 \ifnum #1\expandafter \@firstoftwo
 \else \expandafter \@secondoftwo
 \fi
}%
\providecommand \@ifx [1]{%
 \ifx #1\expandafter \@firstoftwo
 \else \expandafter \@secondoftwo
 \fi
}%
\providecommand \natexlab [1]{#1}%
\providecommand \enquote  [1]{``#1''}%
\providecommand \bibnamefont  [1]{#1}%
\providecommand \bibfnamefont [1]{#1}%
\providecommand \citenamefont [1]{#1}%
\providecommand \href@noop [0]{\@secondoftwo}%
\providecommand \href [0]{\begingroup \@sanitize@url \@href}%
\providecommand \@href[1]{\@@startlink{#1}\@@href}%
\providecommand \@@href[1]{\endgroup#1\@@endlink}%
\providecommand \@sanitize@url [0]{\catcode `\\12\catcode `\$12\catcode `\&12\catcode `\#12\catcode `\^12\catcode `\_12\catcode `\%12\relax}%
\providecommand \@@startlink[1]{}%
\providecommand \@@endlink[0]{}%
\providecommand \url  [0]{\begingroup\@sanitize@url \@url }%
\providecommand \@url [1]{\endgroup\@href {#1}{\urlprefix }}%
\providecommand \urlprefix  [0]{URL }%
\providecommand \Eprint [0]{\href }%
\providecommand \doibase [0]{https://doi.org/}%
\providecommand \selectlanguage [0]{\@gobble}%
\providecommand \bibinfo  [0]{\@secondoftwo}%
\providecommand \bibfield  [0]{\@secondoftwo}%
\providecommand \translation [1]{[#1]}%
\providecommand \BibitemOpen [0]{}%
\providecommand \bibitemStop [0]{}%
\providecommand \bibitemNoStop [0]{.\EOS\space}%
\providecommand \EOS [0]{\spacefactor3000\relax}%
\providecommand \BibitemShut  [1]{\csname bibitem#1\endcsname}%
\let\auto@bib@innerbib\@empty
\bibitem [{\citenamefont {Bohnet}\ \emph {et~al.}(2016)\citenamefont {Bohnet}, \citenamefont {Sawyer}, \citenamefont {Britton}, \citenamefont {Wall}, \citenamefont {Rey}, \citenamefont {{Foss-Feig}},\ and\ \citenamefont {Bollinger}}]{bohnetQuantumSpinDynamics2016}%
  \BibitemOpen
  \bibfield  {author} {\bibinfo {author} {\bibfnamefont {J.~G.}\ \bibnamefont {Bohnet}}, \bibinfo {author} {\bibfnamefont {B.~C.}\ \bibnamefont {Sawyer}}, \bibinfo {author} {\bibfnamefont {J.~W.}\ \bibnamefont {Britton}}, \bibinfo {author} {\bibfnamefont {M.~L.}\ \bibnamefont {Wall}}, \bibinfo {author} {\bibfnamefont {A.~M.}\ \bibnamefont {Rey}}, \bibinfo {author} {\bibfnamefont {M.}~\bibnamefont {{Foss-Feig}}},\ and\ \bibinfo {author} {\bibfnamefont {J.~J.}\ \bibnamefont {Bollinger}},\ }\href {https://doi.org/10.1126/science.aad9958} {\bibfield  {journal} {\bibinfo  {journal} {Science}\ }\textbf {\bibinfo {volume} {352}},\ \bibinfo {pages} {1297} (\bibinfo {year} {2016})}\BibitemShut {NoStop}%
\bibitem [{\citenamefont {Altman}\ \emph {et~al.}(2021)\citenamefont {Altman}, \citenamefont {Brown}, \citenamefont {Carleo}, \citenamefont {Carr}, \citenamefont {Demler}, \citenamefont {Chin}, \citenamefont {DeMarco}, \citenamefont {Economou}, \citenamefont {Eriksson}, \citenamefont {Fu}, \citenamefont {Greiner}, \citenamefont {Hazzard}, \citenamefont {Hulet}, \citenamefont {Koll{\'a}r}, \citenamefont {Lev} \emph {et~al.}}]{altmanQuantumSimulatorsArchitectures2021a}%
  \BibitemOpen
  \bibfield  {author} {\bibinfo {author} {\bibfnamefont {E.}~\bibnamefont {Altman}}, \bibinfo {author} {\bibfnamefont {K.~R.}\ \bibnamefont {Brown}}, \bibinfo {author} {\bibfnamefont {G.}~\bibnamefont {Carleo}}, \bibinfo {author} {\bibfnamefont {L.~D.}\ \bibnamefont {Carr}}, \bibinfo {author} {\bibfnamefont {E.}~\bibnamefont {Demler}}, \bibinfo {author} {\bibfnamefont {C.}~\bibnamefont {Chin}}, \bibinfo {author} {\bibfnamefont {B.}~\bibnamefont {DeMarco}}, \bibinfo {author} {\bibfnamefont {S.~E.}\ \bibnamefont {Economou}}, \bibinfo {author} {\bibfnamefont {M.~A.}\ \bibnamefont {Eriksson}}, \bibinfo {author} {\bibfnamefont {K.-M.~C.}\ \bibnamefont {Fu}}, \bibinfo {author} {\bibfnamefont {M.}~\bibnamefont {Greiner}}, \bibinfo {author} {\bibfnamefont {K.~R.}\ \bibnamefont {Hazzard}}, \bibinfo {author} {\bibfnamefont {R.~G.}\ \bibnamefont {Hulet}}, \bibinfo {author} {\bibfnamefont {A.~J.}\ \bibnamefont {Koll{\'a}r}}, \bibinfo {author} {\bibfnamefont {B.~L.}\ \bibnamefont {Lev}}, \emph {et~al.},\ }\href
  {https://doi.org/10.1103/PRXQuantum.2.017003} {\bibfield  {journal} {\bibinfo  {journal} {PRX Quantum}\ }\textbf {\bibinfo {volume} {2}},\ \bibinfo {pages} {017003} (\bibinfo {year} {2021})}\BibitemShut {NoStop}%
\bibitem [{\citenamefont {Monroe}\ \emph {et~al.}(2021)\citenamefont {Monroe}, \citenamefont {Campbell}, \citenamefont {Duan}, \citenamefont {Gong}, \citenamefont {Gorshkov}, \citenamefont {Hess}, \citenamefont {Islam}, \citenamefont {Kim}, \citenamefont {Linke}, \citenamefont {Pagano}, \citenamefont {Richerme}, \citenamefont {Senko},\ and\ \citenamefont {Yao}}]{monroeProgrammableQuantumSimulations2021}%
  \BibitemOpen
  \bibfield  {author} {\bibinfo {author} {\bibfnamefont {C.}~\bibnamefont {Monroe}}, \bibinfo {author} {\bibfnamefont {W.~C.}\ \bibnamefont {Campbell}}, \bibinfo {author} {\bibfnamefont {L.-M.}\ \bibnamefont {Duan}}, \bibinfo {author} {\bibfnamefont {Z.-X.}\ \bibnamefont {Gong}}, \bibinfo {author} {\bibfnamefont {A.~V.}\ \bibnamefont {Gorshkov}}, \bibinfo {author} {\bibfnamefont {P.~W.}\ \bibnamefont {Hess}}, \bibinfo {author} {\bibfnamefont {R.}~\bibnamefont {Islam}}, \bibinfo {author} {\bibfnamefont {K.}~\bibnamefont {Kim}}, \bibinfo {author} {\bibfnamefont {N.~M.}\ \bibnamefont {Linke}}, \bibinfo {author} {\bibfnamefont {G.}~\bibnamefont {Pagano}}, \bibinfo {author} {\bibfnamefont {P.}~\bibnamefont {Richerme}}, \bibinfo {author} {\bibfnamefont {C.}~\bibnamefont {Senko}},\ and\ \bibinfo {author} {\bibfnamefont {N.~Y.}\ \bibnamefont {Yao}},\ }\href {https://doi.org/10.1103/RevModPhys.93.025001} {\bibfield  {journal} {\bibinfo  {journal} {Rev. Mod. Phys.}\ }\textbf {\bibinfo {volume} {93}},\ \bibinfo {pages}
  {025001} (\bibinfo {year} {2021})}\BibitemShut {NoStop}%
\bibitem [{\citenamefont {Kaer}\ \emph {et~al.}(2010)\citenamefont {Kaer}, \citenamefont {Nielsen}, \citenamefont {Lodahl}, \citenamefont {Jauho},\ and\ \citenamefont {M{\o}rk}}]{kaerNonMarkovianModelPhotonAssisted2010}%
  \BibitemOpen
  \bibfield  {author} {\bibinfo {author} {\bibfnamefont {P.}~\bibnamefont {Kaer}}, \bibinfo {author} {\bibfnamefont {T.~R.}\ \bibnamefont {Nielsen}}, \bibinfo {author} {\bibfnamefont {P.}~\bibnamefont {Lodahl}}, \bibinfo {author} {\bibfnamefont {A.-P.}\ \bibnamefont {Jauho}},\ and\ \bibinfo {author} {\bibfnamefont {J.}~\bibnamefont {M{\o}rk}},\ }\href {https://doi.org/10.1103/PhysRevLett.104.157401} {\bibfield  {journal} {\bibinfo  {journal} {Phys. Rev. Lett.}\ }\textbf {\bibinfo {volume} {104}},\ \bibinfo {pages} {157401} (\bibinfo {year} {2010})}\BibitemShut {NoStop}%
\bibitem [{\citenamefont {Nagy}\ and\ \citenamefont {Domokos}(2015)}]{nagyNonequilibriumQuantumCriticality2015}%
  \BibitemOpen
  \bibfield  {author} {\bibinfo {author} {\bibfnamefont {D.}~\bibnamefont {Nagy}}\ and\ \bibinfo {author} {\bibfnamefont {P.}~\bibnamefont {Domokos}},\ }\href {https://doi.org/10.1103/PhysRevLett.115.043601} {\bibfield  {journal} {\bibinfo  {journal} {Phys. Rev. Lett.}\ }\textbf {\bibinfo {volume} {115}},\ \bibinfo {pages} {043601} (\bibinfo {year} {2015})}\BibitemShut {NoStop}%
\bibitem [{\citenamefont {Cosacchi}\ \emph {et~al.}(2018)\citenamefont {Cosacchi}, \citenamefont {Cygorek}, \citenamefont {Ungar}, \citenamefont {Barth}, \citenamefont {Vagov},\ and\ \citenamefont {Axt}}]{cosacchiPathintegralApproachNonequilibrium2018}%
  \BibitemOpen
  \bibfield  {author} {\bibinfo {author} {\bibfnamefont {M.}~\bibnamefont {Cosacchi}}, \bibinfo {author} {\bibfnamefont {M.}~\bibnamefont {Cygorek}}, \bibinfo {author} {\bibfnamefont {F.}~\bibnamefont {Ungar}}, \bibinfo {author} {\bibfnamefont {A.~M.}\ \bibnamefont {Barth}}, \bibinfo {author} {\bibfnamefont {A.}~\bibnamefont {Vagov}},\ and\ \bibinfo {author} {\bibfnamefont {V.~M.}\ \bibnamefont {Axt}},\ }\href {https://doi.org/10.1103/PhysRevB.98.125302} {\bibfield  {journal} {\bibinfo  {journal} {Phys. Rev. B}\ }\textbf {\bibinfo {volume} {98}},\ \bibinfo {pages} {125302} (\bibinfo {year} {2018})}\BibitemShut {NoStop}%
\bibitem [{\citenamefont {Lundgren}\ \emph {et~al.}(2020)\citenamefont {Lundgren}, \citenamefont {Gorshkov},\ and\ \citenamefont {Maghrebi}}]{lundgrenNatureNonequilibriumPhase2020}%
  \BibitemOpen
  \bibfield  {author} {\bibinfo {author} {\bibfnamefont {R.}~\bibnamefont {Lundgren}}, \bibinfo {author} {\bibfnamefont {A.~V.}\ \bibnamefont {Gorshkov}},\ and\ \bibinfo {author} {\bibfnamefont {M.~F.}\ \bibnamefont {Maghrebi}},\ }\href {https://doi.org/10.1103/PhysRevA.102.032218} {\bibfield  {journal} {\bibinfo  {journal} {Phys. Rev. A}\ }\textbf {\bibinfo {volume} {102}},\ \bibinfo {pages} {032218} (\bibinfo {year} {2020})}\BibitemShut {NoStop}%
\bibitem [{\citenamefont {Lemmer}\ \emph {et~al.}(2018)\citenamefont {Lemmer}, \citenamefont {Cormick}, \citenamefont {Tamascelli}, \citenamefont {Schaetz}, \citenamefont {Huelga},\ and\ \citenamefont {Plenio}}]{lemmerTrappedionSimulatorSpinboson2018}%
  \BibitemOpen
  \bibfield  {author} {\bibinfo {author} {\bibfnamefont {A.}~\bibnamefont {Lemmer}}, \bibinfo {author} {\bibfnamefont {C.}~\bibnamefont {Cormick}}, \bibinfo {author} {\bibfnamefont {D.}~\bibnamefont {Tamascelli}}, \bibinfo {author} {\bibfnamefont {T.}~\bibnamefont {Schaetz}}, \bibinfo {author} {\bibfnamefont {S.~F.}\ \bibnamefont {Huelga}},\ and\ \bibinfo {author} {\bibfnamefont {M.~B.}\ \bibnamefont {Plenio}},\ }\href {https://doi.org/10.1088/1367-2630/aac87d} {\bibfield  {journal} {\bibinfo  {journal} {New J. Phys.}\ }\textbf {\bibinfo {volume} {20}},\ \bibinfo {pages} {073002} (\bibinfo {year} {2018})}\BibitemShut {NoStop}%
\bibitem [{\citenamefont {So}\ \emph {et~al.}(2024)\citenamefont {So}, \citenamefont {Duraisamy~Suganthi}, \citenamefont {Menon}, \citenamefont {Zhu}, \citenamefont {Zhuravel}, \citenamefont {Pu}, \citenamefont {Wolynes}, \citenamefont {Onuchic},\ and\ \citenamefont {Pagano}}]{soTrappedionQuantumSimulation2024}%
  \BibitemOpen
  \bibfield  {author} {\bibinfo {author} {\bibfnamefont {V.}~\bibnamefont {So}}, \bibinfo {author} {\bibfnamefont {M.}~\bibnamefont {Duraisamy~Suganthi}}, \bibinfo {author} {\bibfnamefont {A.}~\bibnamefont {Menon}}, \bibinfo {author} {\bibfnamefont {M.}~\bibnamefont {Zhu}}, \bibinfo {author} {\bibfnamefont {R.}~\bibnamefont {Zhuravel}}, \bibinfo {author} {\bibfnamefont {H.}~\bibnamefont {Pu}}, \bibinfo {author} {\bibfnamefont {P.~G.}\ \bibnamefont {Wolynes}}, \bibinfo {author} {\bibfnamefont {J.~N.}\ \bibnamefont {Onuchic}},\ and\ \bibinfo {author} {\bibfnamefont {G.}~\bibnamefont {Pagano}},\ }\href {https://doi.org/10.1126/sciadv.ads8011} {\bibfield  {journal} {\bibinfo  {journal} {Sci. Adv.}\ }\textbf {\bibinfo {volume} {10}},\ \bibinfo {pages} {eads8011} (\bibinfo {year} {2024})}\BibitemShut {NoStop}%
\bibitem [{\citenamefont {So}\ \emph {et~al.}(2026)\citenamefont {So}, \citenamefont {Duraisamy~Suganthi}, \citenamefont {Zhu}, \citenamefont {Menon}, \citenamefont {Tomaras}, \citenamefont {Zhuravel}, \citenamefont {Pu}, \citenamefont {Wolynes}, \citenamefont {Onuchic},\ and\ \citenamefont {Pagano}}]{soQuantumSimulationCharge2025}%
  \BibitemOpen
  \bibfield  {author} {\bibinfo {author} {\bibfnamefont {V.}~\bibnamefont {So}}, \bibinfo {author} {\bibfnamefont {M.}~\bibnamefont {Duraisamy~Suganthi}}, \bibinfo {author} {\bibfnamefont {M.}~\bibnamefont {Zhu}}, \bibinfo {author} {\bibfnamefont {A.}~\bibnamefont {Menon}}, \bibinfo {author} {\bibfnamefont {G.}~\bibnamefont {Tomaras}}, \bibinfo {author} {\bibfnamefont {R.}~\bibnamefont {Zhuravel}}, \bibinfo {author} {\bibfnamefont {H.}~\bibnamefont {Pu}}, \bibinfo {author} {\bibfnamefont {P.~G.}\ \bibnamefont {Wolynes}}, \bibinfo {author} {\bibfnamefont {J.~N.}\ \bibnamefont {Onuchic}},\ and\ \bibinfo {author} {\bibfnamefont {G.}~\bibnamefont {Pagano}},\ }\href {https://doi.org/10.1038/s41467-025-67116-6} {\bibfield  {journal} {\bibinfo  {journal} {Nat Commun}\ }\textbf {\bibinfo {volume} {17}},\ \bibinfo {pages} {438} (\bibinfo {year} {2026})}\BibitemShut {NoStop}%
\bibitem [{\citenamefont {Kamar}\ \emph {et~al.}(2024)\citenamefont {Kamar}, \citenamefont {Paz},\ and\ \citenamefont {Maghrebi}}]{kamarSpinbosonModelDephasing2024}%
  \BibitemOpen
  \bibfield  {author} {\bibinfo {author} {\bibfnamefont {N.~A.}\ \bibnamefont {Kamar}}, \bibinfo {author} {\bibfnamefont {D.~A.}\ \bibnamefont {Paz}},\ and\ \bibinfo {author} {\bibfnamefont {M.~F.}\ \bibnamefont {Maghrebi}},\ }\href {https://doi.org/10.1103/PhysRevB.110.075126} {\bibfield  {journal} {\bibinfo  {journal} {Phys. Rev. B}\ }\textbf {\bibinfo {volume} {110}},\ \bibinfo {pages} {075126} (\bibinfo {year} {2024})}\BibitemShut {NoStop}%
\bibitem [{\citenamefont {Sun}\ \emph {et~al.}(2025)\citenamefont {Sun}, \citenamefont {Kang}, \citenamefont {Nuomin}, \citenamefont {Schwartz}, \citenamefont {Beratan}, \citenamefont {Brown},\ and\ \citenamefont {Kim}}]{sunQuantumSimulationSpinboson2025a}%
  \BibitemOpen
  \bibfield  {author} {\bibinfo {author} {\bibfnamefont {K.}~\bibnamefont {Sun}}, \bibinfo {author} {\bibfnamefont {M.}~\bibnamefont {Kang}}, \bibinfo {author} {\bibfnamefont {H.}~\bibnamefont {Nuomin}}, \bibinfo {author} {\bibfnamefont {G.}~\bibnamefont {Schwartz}}, \bibinfo {author} {\bibfnamefont {D.~N.}\ \bibnamefont {Beratan}}, \bibinfo {author} {\bibfnamefont {K.~R.}\ \bibnamefont {Brown}},\ and\ \bibinfo {author} {\bibfnamefont {J.}~\bibnamefont {Kim}},\ }\href {https://doi.org/10.1038/s41467-025-59296-y} {\bibfield  {journal} {\bibinfo  {journal} {Nat Commun}\ }\textbf {\bibinfo {volume} {16}},\ \bibinfo {pages} {4042} (\bibinfo {year} {2025})}\BibitemShut {NoStop}%
\bibitem [{\citenamefont {De~Vega}\ and\ \citenamefont {Alonso}(2017)}]{devegaDynamicsNonMarkovianOpen2017}%
  \BibitemOpen
  \bibfield  {author} {\bibinfo {author} {\bibfnamefont {I.}~\bibnamefont {De~Vega}}\ and\ \bibinfo {author} {\bibfnamefont {D.}~\bibnamefont {Alonso}},\ }\href {https://doi.org/10.1103/RevModPhys.89.015001} {\bibfield  {journal} {\bibinfo  {journal} {Rev. Mod. Phys.}\ }\textbf {\bibinfo {volume} {89}},\ \bibinfo {pages} {015001} (\bibinfo {year} {2017})}\BibitemShut {NoStop}%
\bibitem [{\citenamefont {Xu}\ \emph {et~al.}(2026)\citenamefont {Xu}, \citenamefont {Vadimov}, \citenamefont {Stockburger},\ and\ \citenamefont {Ankerhold}}]{xuColloquiumSimulatingNonMarkovian2026}%
  \BibitemOpen
  \bibfield  {author} {\bibinfo {author} {\bibfnamefont {M.}~\bibnamefont {Xu}}, \bibinfo {author} {\bibfnamefont {V.}~\bibnamefont {Vadimov}}, \bibinfo {author} {\bibfnamefont {J.~T.}\ \bibnamefont {Stockburger}},\ and\ \bibinfo {author} {\bibfnamefont {J.}~\bibnamefont {Ankerhold}},\ }\href {https://doi.org/10.1103/w3nw-hbjc} {\bibfield  {journal} {\bibinfo  {journal} {Rev. Mod. Phys.}\ }\textbf {\bibinfo {volume} {98}},\ \bibinfo {pages} {021002} (\bibinfo {year} {2026})}\BibitemShut {NoStop}%
\bibitem [{\citenamefont {Keeling}\ \emph {et~al.}(2026)\citenamefont {Keeling}, \citenamefont {Stoudenmire}, \citenamefont {Ba{\~n}uls},\ and\ \citenamefont {Reichman}}]{keelingProcessTensorApproaches2026}%
  \BibitemOpen
  \bibfield  {author} {\bibinfo {author} {\bibfnamefont {J.}~\bibnamefont {Keeling}}, \bibinfo {author} {\bibfnamefont {E.~M.}\ \bibnamefont {Stoudenmire}}, \bibinfo {author} {\bibfnamefont {M.-C.}\ \bibnamefont {Ba{\~n}uls}},\ and\ \bibinfo {author} {\bibfnamefont {D.~R.}\ \bibnamefont {Reichman}},\ }\href {https://doi.org/10.1103/1ncg-11hz} {\bibfield  {journal} {\bibinfo  {journal} {Phys. Rev. X}\ }\textbf {\bibinfo {volume} {16}},\ \bibinfo {pages} {020502} (\bibinfo {year} {2026})}\BibitemShut {NoStop}%
\bibitem [{\citenamefont {Gandhari}\ and\ \citenamefont {Gullans}(2026)}]{gandhariQuantumNonMarkovianNoise2026a}%
  \BibitemOpen
  \bibfield  {author} {\bibinfo {author} {\bibfnamefont {S.}~\bibnamefont {Gandhari}}\ and\ \bibinfo {author} {\bibfnamefont {M.~J.}\ \bibnamefont {Gullans}},\ }\href {https://doi.org/10.1103/z8zh-n1hl} {\bibfield  {journal} {\bibinfo  {journal} {Phys. Rev. Research}\ }\textbf {\bibinfo {volume} {8}},\ \bibinfo {pages} {023075} (\bibinfo {year} {2026})}\BibitemShut {NoStop}%
\bibitem [{\citenamefont {Betzholz}\ \emph {et~al.}(2014)\citenamefont {Betzholz}, \citenamefont {Torres},\ and\ \citenamefont {Bienert}}]{betzholzQuantumOpticalMaster2014}%
  \BibitemOpen
  \bibfield  {author} {\bibinfo {author} {\bibfnamefont {R.}~\bibnamefont {Betzholz}}, \bibinfo {author} {\bibfnamefont {J.~M.}\ \bibnamefont {Torres}},\ and\ \bibinfo {author} {\bibfnamefont {M.}~\bibnamefont {Bienert}},\ }\href {https://doi.org/10.1103/PhysRevA.90.063818} {\bibfield  {journal} {\bibinfo  {journal} {Phys. Rev. A}\ }\textbf {\bibinfo {volume} {90}},\ \bibinfo {pages} {063818} (\bibinfo {year} {2014})}\BibitemShut {NoStop}%
\bibitem [{\citenamefont {Johnsson}\ \emph {et~al.}(2022)\citenamefont {Johnsson}, \citenamefont {Baragiola}, \citenamefont {Volz},\ and\ \citenamefont {Brennen}}]{johnssonModifiedCoherenceQuantum2022}%
  \BibitemOpen
  \bibfield  {author} {\bibinfo {author} {\bibfnamefont {M.~T.}\ \bibnamefont {Johnsson}}, \bibinfo {author} {\bibfnamefont {B.~Q.}\ \bibnamefont {Baragiola}}, \bibinfo {author} {\bibfnamefont {T.}~\bibnamefont {Volz}},\ and\ \bibinfo {author} {\bibfnamefont {G.~K.}\ \bibnamefont {Brennen}},\ }\href {https://doi.org/10.1103/PhysRevB.105.094308} {\bibfield  {journal} {\bibinfo  {journal} {Phys. Rev. B}\ }\textbf {\bibinfo {volume} {105}},\ \bibinfo {pages} {094308} (\bibinfo {year} {2022})}\BibitemShut {NoStop}%
\bibitem [{\citenamefont {Kamar}\ and\ \citenamefont {Maghrebi}(2025)}]{kamarHybridQuantumClassicalStochastic2025a}%
  \BibitemOpen
  \bibfield  {author} {\bibinfo {author} {\bibfnamefont {N.~A.}\ \bibnamefont {Kamar}}\ and\ \bibinfo {author} {\bibfnamefont {M.}~\bibnamefont {Maghrebi}},\ }\href {https://doi.org/10.1103/PhysRevX.15.021073} {\bibfield  {journal} {\bibinfo  {journal} {Phys. Rev. X}\ }\textbf {\bibinfo {volume} {15}},\ \bibinfo {pages} {021073} (\bibinfo {year} {2025})}\BibitemShut {NoStop}%
\bibitem [{\citenamefont {Dimer}\ \emph {et~al.}(2007)\citenamefont {Dimer}, \citenamefont {Estienne}, \citenamefont {Parkins},\ and\ \citenamefont {Carmichael}}]{dimerProposedRealizationDickemodel2007}%
  \BibitemOpen
  \bibfield  {author} {\bibinfo {author} {\bibfnamefont {F.}~\bibnamefont {Dimer}}, \bibinfo {author} {\bibfnamefont {B.}~\bibnamefont {Estienne}}, \bibinfo {author} {\bibfnamefont {A.~S.}\ \bibnamefont {Parkins}},\ and\ \bibinfo {author} {\bibfnamefont {H.~J.}\ \bibnamefont {Carmichael}},\ }\href {https://doi.org/10.1103/PhysRevA.75.013804} {\bibfield  {journal} {\bibinfo  {journal} {Phys. Rev. A}\ }\textbf {\bibinfo {volume} {75}},\ \bibinfo {pages} {013804} (\bibinfo {year} {2007})}\BibitemShut {NoStop}%
\bibitem [{\citenamefont {Metelmann}\ and\ \citenamefont {Clerk}(2015)}]{metelmannNonreciprocalPhotonTransmission2015a}%
  \BibitemOpen
  \bibfield  {author} {\bibinfo {author} {\bibfnamefont {A.}~\bibnamefont {Metelmann}}\ and\ \bibinfo {author} {\bibfnamefont {A.~A.}\ \bibnamefont {Clerk}},\ }\href {https://doi.org/10.1103/PhysRevX.5.021025} {\bibfield  {journal} {\bibinfo  {journal} {Phys. Rev. X}\ }\textbf {\bibinfo {volume} {5}},\ \bibinfo {pages} {021025} (\bibinfo {year} {2015})}\BibitemShut {NoStop}%
\bibitem [{\citenamefont {Seif}\ \emph {et~al.}(2022)\citenamefont {Seif}, \citenamefont {Wang},\ and\ \citenamefont {Clerk}}]{seifDistinguishingQuantumClassical2022}%
  \BibitemOpen
  \bibfield  {author} {\bibinfo {author} {\bibfnamefont {A.}~\bibnamefont {Seif}}, \bibinfo {author} {\bibfnamefont {Y.-X.}\ \bibnamefont {Wang}},\ and\ \bibinfo {author} {\bibfnamefont {A.~A.}\ \bibnamefont {Clerk}},\ }\href {https://doi.org/10.1103/PhysRevLett.128.070402} {\bibfield  {journal} {\bibinfo  {journal} {Phys. Rev. Lett.}\ }\textbf {\bibinfo {volume} {128}},\ \bibinfo {pages} {070402} (\bibinfo {year} {2022})}\BibitemShut {NoStop}%
\bibitem [{\citenamefont {Stockburger}\ and\ \citenamefont {Grabert}(2002)}]{stockburgerExactNumberRepresentation2002}%
  \BibitemOpen
  \bibfield  {author} {\bibinfo {author} {\bibfnamefont {J.~T.}\ \bibnamefont {Stockburger}}\ and\ \bibinfo {author} {\bibfnamefont {H.}~\bibnamefont {Grabert}},\ }\href {https://doi.org/10.1103/PhysRevLett.88.170407} {\bibfield  {journal} {\bibinfo  {journal} {Phys. Rev. Lett.}\ }\textbf {\bibinfo {volume} {88}},\ \bibinfo {pages} {170407} (\bibinfo {year} {2002})}\BibitemShut {NoStop}%
\bibitem [{sm()}]{sm}%
  \BibitemOpen
  \href@noop {} {}\bibinfo {note} {See Supplemental Material for (I) a proof of the exact stochastic representation; (II) spin and bosonic observables without local spin dissipation; (III) exact solutions with dephasing, relaxation, and pumping; (IV) coherence-generating dissipation; (V) the spin-only time-local equation, adiabatic elimination, and subsystem dynamics with nonreciprocal interactions; and (VI) details of the spin-squeezing calculation. The Supplemental Material includes Refs.~\cite{polkovnikovPhaseSpaceRepresentation2010, horodeckiGeneralEntanglementBreaking2003, seifCompressedSensingMeasurement2021a, pezzeQuantumMetrologyNonclassical2018}.}\BibitemShut {Stop}%
\bibitem [{\citenamefont {McDonald}\ and\ \citenamefont {Clerk}(2022)}]{mcdonaldExactSolutionsInteracting2022}%
  \BibitemOpen
  \bibfield  {author} {\bibinfo {author} {\bibfnamefont {A.}~\bibnamefont {McDonald}}\ and\ \bibinfo {author} {\bibfnamefont {A.~A.}\ \bibnamefont {Clerk}},\ }\href {https://doi.org/10.1103/PhysRevLett.128.033602} {\bibfield  {journal} {\bibinfo  {journal} {Phys. Rev. Lett.}\ }\textbf {\bibinfo {volume} {128}},\ \bibinfo {pages} {033602} (\bibinfo {year} {2022})}\BibitemShut {NoStop}%
\bibitem [{\citenamefont {Hartmann}\ \emph {et~al.}(2006)\citenamefont {Hartmann}, \citenamefont {D{\"u}r},\ and\ \citenamefont {Briegel}}]{hartmannSteadystateEntanglementOpen2006}%
  \BibitemOpen
  \bibfield  {author} {\bibinfo {author} {\bibfnamefont {L.}~\bibnamefont {Hartmann}}, \bibinfo {author} {\bibfnamefont {W.}~\bibnamefont {D{\"u}r}},\ and\ \bibinfo {author} {\bibfnamefont {H.-J.}\ \bibnamefont {Briegel}},\ }\href {https://doi.org/10.1103/PhysRevA.74.052304} {\bibfield  {journal} {\bibinfo  {journal} {Phys. Rev. A}\ }\textbf {\bibinfo {volume} {74}},\ \bibinfo {pages} {052304} (\bibinfo {year} {2006})}\BibitemShut {NoStop}%
\bibitem [{\citenamefont {{Foss-Feig}}\ \emph {et~al.}(2017)\citenamefont {{Foss-Feig}}, \citenamefont {Young}, \citenamefont {Albert}, \citenamefont {Gorshkov},\ and\ \citenamefont {Maghrebi}}]{foss-feigSolvableFamilyDrivenDissipative2017}%
  \BibitemOpen
  \bibfield  {author} {\bibinfo {author} {\bibfnamefont {M.}~\bibnamefont {{Foss-Feig}}}, \bibinfo {author} {\bibfnamefont {J.~T.}\ \bibnamefont {Young}}, \bibinfo {author} {\bibfnamefont {V.~V.}\ \bibnamefont {Albert}}, \bibinfo {author} {\bibfnamefont {A.~V.}\ \bibnamefont {Gorshkov}},\ and\ \bibinfo {author} {\bibfnamefont {M.~F.}\ \bibnamefont {Maghrebi}},\ }\href {https://doi.org/10.1103/PhysRevLett.119.190402} {\bibfield  {journal} {\bibinfo  {journal} {Phys. Rev. Lett.}\ }\textbf {\bibinfo {volume} {119}},\ \bibinfo {pages} {190402} (\bibinfo {year} {2017})}\BibitemShut {NoStop}%
\bibitem [{\citenamefont {Zou}\ \emph {et~al.}(2024)\citenamefont {Zou}, \citenamefont {Bosco},\ and\ \citenamefont {Loss}}]{zouSpatiallyCorrelatedClassical2024b}%
  \BibitemOpen
  \bibfield  {author} {\bibinfo {author} {\bibfnamefont {J.}~\bibnamefont {Zou}}, \bibinfo {author} {\bibfnamefont {S.}~\bibnamefont {Bosco}},\ and\ \bibinfo {author} {\bibfnamefont {D.}~\bibnamefont {Loss}},\ }\href {https://doi.org/10.1038/s41534-024-00842-9} {\bibfield  {journal} {\bibinfo  {journal} {npj Quantum Inf}\ }\textbf {\bibinfo {volume} {10}},\ \bibinfo {pages} {46} (\bibinfo {year} {2024})}\BibitemShut {NoStop}%
\bibitem [{\citenamefont {Jachinowski}\ and\ \citenamefont {Littlewood}(2026)}]{jachinowskiSpinonlyDynamicsMultispecies2026}%
  \BibitemOpen
  \bibfield  {author} {\bibinfo {author} {\bibfnamefont {J.}~\bibnamefont {Jachinowski}}\ and\ \bibinfo {author} {\bibfnamefont {P.~B.}\ \bibnamefont {Littlewood}},\ }\href {https://doi.org/10.21468/SciPostPhys.20.6.155} {\bibfield  {journal} {\bibinfo  {journal} {SciPost Phys.}\ }\textbf {\bibinfo {volume} {20}},\ \bibinfo {pages} {155} (\bibinfo {year} {2026})}\BibitemShut {NoStop}%
\bibitem [{\citenamefont {{Foss-Feig}}\ \emph {et~al.}(2013{\natexlab{a}})\citenamefont {{Foss-Feig}}, \citenamefont {Hazzard}, \citenamefont {Bollinger},\ and\ \citenamefont {Rey}}]{foss-feigNonequilibriumDynamicsArbitraryrange2013a}%
  \BibitemOpen
  \bibfield  {author} {\bibinfo {author} {\bibfnamefont {M.}~\bibnamefont {{Foss-Feig}}}, \bibinfo {author} {\bibfnamefont {K.~R.~A.}\ \bibnamefont {Hazzard}}, \bibinfo {author} {\bibfnamefont {J.~J.}\ \bibnamefont {Bollinger}},\ and\ \bibinfo {author} {\bibfnamefont {A.~M.}\ \bibnamefont {Rey}},\ }\href {https://doi.org/10.1103/PhysRevA.87.042101} {\bibfield  {journal} {\bibinfo  {journal} {Phys. Rev. A}\ }\textbf {\bibinfo {volume} {87}},\ \bibinfo {pages} {042101} (\bibinfo {year} {2013}{\natexlab{a}})}\BibitemShut {NoStop}%
\bibitem [{\citenamefont {{Foss-Feig}}\ \emph {et~al.}(2013{\natexlab{b}})\citenamefont {{Foss-Feig}}, \citenamefont {Hazzard}, \citenamefont {Bollinger}, \citenamefont {Rey},\ and\ \citenamefont {Clark}}]{foss-feigDynamicalQuantumCorrelations2013}%
  \BibitemOpen
  \bibfield  {author} {\bibinfo {author} {\bibfnamefont {M.}~\bibnamefont {{Foss-Feig}}}, \bibinfo {author} {\bibfnamefont {K.~R.~A.}\ \bibnamefont {Hazzard}}, \bibinfo {author} {\bibfnamefont {J.~J.}\ \bibnamefont {Bollinger}}, \bibinfo {author} {\bibfnamefont {A.~M.}\ \bibnamefont {Rey}},\ and\ \bibinfo {author} {\bibfnamefont {C.~W.}\ \bibnamefont {Clark}},\ }\href {https://doi.org/10.1088/1367-2630/15/11/113008} {\bibfield  {journal} {\bibinfo  {journal} {New J. Phys.}\ }\textbf {\bibinfo {volume} {15}},\ \bibinfo {pages} {113008} (\bibinfo {year} {2013}{\natexlab{b}})}\BibitemShut {NoStop}%
\bibitem [{\citenamefont {Dylewsky}\ \emph {et~al.}(2016)\citenamefont {Dylewsky}, \citenamefont {Freericks}, \citenamefont {Wall}, \citenamefont {Rey},\ and\ \citenamefont {{Foss-Feig}}}]{dylewskyNonperturbativeCalculationPhonon2016}%
  \BibitemOpen
  \bibfield  {author} {\bibinfo {author} {\bibfnamefont {D.}~\bibnamefont {Dylewsky}}, \bibinfo {author} {\bibfnamefont {J.~K.}\ \bibnamefont {Freericks}}, \bibinfo {author} {\bibfnamefont {M.~L.}\ \bibnamefont {Wall}}, \bibinfo {author} {\bibfnamefont {A.~M.}\ \bibnamefont {Rey}},\ and\ \bibinfo {author} {\bibfnamefont {M.}~\bibnamefont {{Foss-Feig}}},\ }\href {https://doi.org/10.1103/PhysRevA.93.013415} {\bibfield  {journal} {\bibinfo  {journal} {Phys. Rev. A}\ }\textbf {\bibinfo {volume} {93}},\ \bibinfo {pages} {013415} (\bibinfo {year} {2016})}\BibitemShut {NoStop}%
\bibitem [{\citenamefont {Wineland}\ \emph {et~al.}(1994)\citenamefont {Wineland}, \citenamefont {Bollinger}, \citenamefont {Itano},\ and\ \citenamefont {Heinzen}}]{winelandSqueezedAtomicStates1994}%
  \BibitemOpen
  \bibfield  {author} {\bibinfo {author} {\bibfnamefont {D.~J.}\ \bibnamefont {Wineland}}, \bibinfo {author} {\bibfnamefont {J.~J.}\ \bibnamefont {Bollinger}}, \bibinfo {author} {\bibfnamefont {W.~M.}\ \bibnamefont {Itano}},\ and\ \bibinfo {author} {\bibfnamefont {D.~J.}\ \bibnamefont {Heinzen}},\ }\href {https://doi.org/10.1103/PhysRevA.50.67} {\bibfield  {journal} {\bibinfo  {journal} {Phys. Rev. A}\ }\textbf {\bibinfo {volume} {50}},\ \bibinfo {pages} {67} (\bibinfo {year} {1994})}\BibitemShut {NoStop}%
\bibitem [{\citenamefont {Torres}\ \emph {et~al.}(2019)\citenamefont {Torres}, \citenamefont {Betzholz},\ and\ \citenamefont {Bienert}}]{torresOptomechanicalDampingBasis2019}%
  \BibitemOpen
  \bibfield  {author} {\bibinfo {author} {\bibfnamefont {J.~M.}\ \bibnamefont {Torres}}, \bibinfo {author} {\bibfnamefont {R.}~\bibnamefont {Betzholz}},\ and\ \bibinfo {author} {\bibfnamefont {M.}~\bibnamefont {Bienert}},\ }\href {https://doi.org/10.1088/1751-8121/aafffe} {\bibfield  {journal} {\bibinfo  {journal} {J. Phys. A: Math. Theor.}\ }\textbf {\bibinfo {volume} {52}},\ \bibinfo {pages} {08LT02} (\bibinfo {year} {2019})}\BibitemShut {NoStop}%
\bibitem [{\citenamefont {Polkovnikov}(2010)}]{polkovnikovPhaseSpaceRepresentation2010}%
  \BibitemOpen
  \bibfield  {author} {\bibinfo {author} {\bibfnamefont {A.}~\bibnamefont {Polkovnikov}},\ }\href {https://doi.org/10.1016/j.aop.2010.02.006} {\bibfield  {journal} {\bibinfo  {journal} {Annals of Physics}\ }\textbf {\bibinfo {volume} {325}},\ \bibinfo {pages} {1790} (\bibinfo {year} {2010})}\BibitemShut {NoStop}%
\bibitem [{\citenamefont {Horodecki}\ \emph {et~al.}(2003)\citenamefont {Horodecki}, \citenamefont {Shor},\ and\ \citenamefont {Ruskai}}]{horodeckiGeneralEntanglementBreaking2003}%
  \BibitemOpen
  \bibfield  {author} {\bibinfo {author} {\bibfnamefont {M.}~\bibnamefont {Horodecki}}, \bibinfo {author} {\bibfnamefont {P.~W.}\ \bibnamefont {Shor}},\ and\ \bibinfo {author} {\bibfnamefont {M.~B.}\ \bibnamefont {Ruskai}},\ }\href {https://doi.org/10.1142/S0129055X03001709} {\bibfield  {journal} {\bibinfo  {journal} {Rev. Math. Phys.}\ }\textbf {\bibinfo {volume} {15}},\ \bibinfo {pages} {629} (\bibinfo {year} {2003})}\BibitemShut {NoStop}%
\bibitem [{\citenamefont {Seif}\ \emph {et~al.}(2021)\citenamefont {Seif}, \citenamefont {Hafezi},\ and\ \citenamefont {Liu}}]{seifCompressedSensingMeasurement2021a}%
  \BibitemOpen
  \bibfield  {author} {\bibinfo {author} {\bibfnamefont {A.}~\bibnamefont {Seif}}, \bibinfo {author} {\bibfnamefont {M.}~\bibnamefont {Hafezi}},\ and\ \bibinfo {author} {\bibfnamefont {Y.-K.}\ \bibnamefont {Liu}},\ }\href@noop {} {\bibinfo {title} {Compressed sensing measurement of long-range correlated noise}} (\bibinfo {year} {2021}),\ \Eprint {https://arxiv.org/abs/2105.12589} {arXiv:2105.12589 [quant-ph]} \BibitemShut {NoStop}%
\bibitem [{\citenamefont {Pezz{\`e}}\ \emph {et~al.}(2018)\citenamefont {Pezz{\`e}}, \citenamefont {Smerzi}, \citenamefont {Oberthaler}, \citenamefont {Schmied},\ and\ \citenamefont {Treutlein}}]{pezzeQuantumMetrologyNonclassical2018}%
  \BibitemOpen
  \bibfield  {author} {\bibinfo {author} {\bibfnamefont {L.}~\bibnamefont {Pezz{\`e}}}, \bibinfo {author} {\bibfnamefont {A.}~\bibnamefont {Smerzi}}, \bibinfo {author} {\bibfnamefont {M.~K.}\ \bibnamefont {Oberthaler}}, \bibinfo {author} {\bibfnamefont {R.}~\bibnamefont {Schmied}},\ and\ \bibinfo {author} {\bibfnamefont {P.}~\bibnamefont {Treutlein}},\ }\href {https://doi.org/10.1103/RevModPhys.90.035005} {\bibfield  {journal} {\bibinfo  {journal} {Rev. Mod. Phys.}\ }\textbf {\bibinfo {volume} {90}},\ \bibinfo {pages} {035005} (\bibinfo {year} {2018})}\BibitemShut {NoStop}%
\end{thebibliography}%


\begin{thebibliography}{10}%
\makeatletter
\providecommand \@ifxundefined [1]{%
 \@ifx{#1\undefined}
}%
\providecommand \@ifnum [1]{%
 \ifnum #1\expandafter \@firstoftwo
 \else \expandafter \@secondoftwo
 \fi
}%
\providecommand \@ifx [1]{%
 \ifx #1\expandafter \@firstoftwo
 \else \expandafter \@secondoftwo
 \fi
}%
\providecommand \natexlab [1]{#1}%
\providecommand \enquote  [1]{``#1''}%
\providecommand \bibnamefont  [1]{#1}%
\providecommand \bibfnamefont [1]{#1}%
\providecommand \citenamefont [1]{#1}%
\providecommand \href@noop [0]{\@secondoftwo}%
\providecommand \href [0]{\begingroup \@sanitize@url \@href}%
\providecommand \@href[1]{\@@startlink{#1}\@@href}%
\providecommand \@@href[1]{\endgroup#1\@@endlink}%
\providecommand \@sanitize@url [0]{\catcode `\\12\catcode `\$12\catcode `\&12\catcode `\#12\catcode `\^12\catcode `\_12\catcode `\%12\relax}%
\providecommand \@@startlink[1]{}%
\providecommand \@@endlink[0]{}%
\providecommand \url  [0]{\begingroup\@sanitize@url \@url }%
\providecommand \@url [1]{\endgroup\@href {#1}{\urlprefix }}%
\providecommand \urlprefix  [0]{URL }%
\providecommand \Eprint [0]{\href }%
\providecommand \doibase [0]{https://doi.org/}%
\providecommand \selectlanguage [0]{\@gobble}%
\providecommand \bibinfo  [0]{\@secondoftwo}%
\providecommand \bibfield  [0]{\@secondoftwo}%
\providecommand \translation [1]{[#1]}%
\providecommand \BibitemOpen [0]{}%
\providecommand \bibitemStop [0]{}%
\providecommand \bibitemNoStop [0]{.\EOS\space}%
\providecommand \EOS [0]{\spacefactor3000\relax}%
\providecommand \BibitemShut  [1]{\csname bibitem#1\endcsname}%
\let\auto@bib@innerbib\@empty
\bibitem [{\citenamefont {Kamar}\ and\ \citenamefont {Maghrebi}(2025)}]{kamarHybridQuantumClassicalStochastic2025a}%
  \BibitemOpen
  \bibfield  {author} {\bibinfo {author} {\bibfnamefont {N.~A.}\ \bibnamefont {Kamar}}\ and\ \bibinfo {author} {\bibfnamefont {M.}~\bibnamefont {Maghrebi}},\ }\href {https://doi.org/10.1103/PhysRevX.15.021073} {\bibfield  {journal} {\bibinfo  {journal} {Phys. Rev. X}\ }\textbf {\bibinfo {volume} {15}},\ \bibinfo {pages} {021073} (\bibinfo {year} {2025})}\BibitemShut {NoStop}%
\bibitem [{\citenamefont {Polkovnikov}(2010)}]{polkovnikovPhaseSpaceRepresentation2010}%
  \BibitemOpen
  \bibfield  {author} {\bibinfo {author} {\bibfnamefont {A.}~\bibnamefont {Polkovnikov}},\ }\href {https://doi.org/10.1016/j.aop.2010.02.006} {\bibfield  {journal} {\bibinfo  {journal} {Annals of Physics}\ }\textbf {\bibinfo {volume} {325}},\ \bibinfo {pages} {1790} (\bibinfo {year} {2010})}\BibitemShut {NoStop}%
\bibitem [{\citenamefont {Horodecki}\ \emph {et~al.}(2003)\citenamefont {Horodecki}, \citenamefont {Shor},\ and\ \citenamefont {Ruskai}}]{horodeckiGeneralEntanglementBreaking2003}%
  \BibitemOpen
  \bibfield  {author} {\bibinfo {author} {\bibfnamefont {M.}~\bibnamefont {Horodecki}}, \bibinfo {author} {\bibfnamefont {P.~W.}\ \bibnamefont {Shor}},\ and\ \bibinfo {author} {\bibfnamefont {M.~B.}\ \bibnamefont {Ruskai}},\ }\href {https://doi.org/10.1142/S0129055X03001709} {\bibfield  {journal} {\bibinfo  {journal} {Rev. Math. Phys.}\ }\textbf {\bibinfo {volume} {15}},\ \bibinfo {pages} {629} (\bibinfo {year} {2003})}\BibitemShut {NoStop}%
\bibitem [{\citenamefont {De~Vega}\ and\ \citenamefont {Alonso}(2017)}]{devegaDynamicsNonMarkovianOpen2017}%
  \BibitemOpen
  \bibfield  {author} {\bibinfo {author} {\bibfnamefont {I.}~\bibnamefont {De~Vega}}\ and\ \bibinfo {author} {\bibfnamefont {D.}~\bibnamefont {Alonso}},\ }\href {https://doi.org/10.1103/RevModPhys.89.015001} {\bibfield  {journal} {\bibinfo  {journal} {Rev. Mod. Phys.}\ }\textbf {\bibinfo {volume} {89}},\ \bibinfo {pages} {015001} (\bibinfo {year} {2017})}\BibitemShut {NoStop}%
\bibitem [{\citenamefont {{Foss-Feig}}\ \emph {et~al.}(2013)\citenamefont {{Foss-Feig}}, \citenamefont {Hazzard}, \citenamefont {Bollinger},\ and\ \citenamefont {Rey}}]{foss-feigNonequilibriumDynamicsArbitraryrange2013a}%
  \BibitemOpen
  \bibfield  {author} {\bibinfo {author} {\bibfnamefont {M.}~\bibnamefont {{Foss-Feig}}}, \bibinfo {author} {\bibfnamefont {K.~R.~A.}\ \bibnamefont {Hazzard}}, \bibinfo {author} {\bibfnamefont {J.~J.}\ \bibnamefont {Bollinger}},\ and\ \bibinfo {author} {\bibfnamefont {A.~M.}\ \bibnamefont {Rey}},\ }\href {https://doi.org/10.1103/PhysRevA.87.042101} {\bibfield  {journal} {\bibinfo  {journal} {Phys. Rev. A}\ }\textbf {\bibinfo {volume} {87}},\ \bibinfo {pages} {042101} (\bibinfo {year} {2013})}\BibitemShut {NoStop}%
\bibitem [{\citenamefont {McDonald}\ and\ \citenamefont {Clerk}(2022)}]{mcdonaldExactSolutionsInteracting2022}%
  \BibitemOpen
  \bibfield  {author} {\bibinfo {author} {\bibfnamefont {A.}~\bibnamefont {McDonald}}\ and\ \bibinfo {author} {\bibfnamefont {A.~A.}\ \bibnamefont {Clerk}},\ }\href {https://doi.org/10.1103/PhysRevLett.128.033602} {\bibfield  {journal} {\bibinfo  {journal} {Phys. Rev. Lett.}\ }\textbf {\bibinfo {volume} {128}},\ \bibinfo {pages} {033602} (\bibinfo {year} {2022})}\BibitemShut {NoStop}%
\bibitem [{\citenamefont {Seif}\ \emph {et~al.}(2021)\citenamefont {Seif}, \citenamefont {Hafezi},\ and\ \citenamefont {Liu}}]{seifCompressedSensingMeasurement2021a}%
  \BibitemOpen
  \bibfield  {author} {\bibinfo {author} {\bibfnamefont {A.}~\bibnamefont {Seif}}, \bibinfo {author} {\bibfnamefont {M.}~\bibnamefont {Hafezi}},\ and\ \bibinfo {author} {\bibfnamefont {Y.-K.}\ \bibnamefont {Liu}},\ }\href@noop {} {\bibinfo {title} {Compressed sensing measurement of long-range correlated noise}} (\bibinfo {year} {2021}),\ \Eprint {https://arxiv.org/abs/2105.12589} {arXiv:2105.12589 [quant-ph]} \BibitemShut {NoStop}%
\bibitem [{\citenamefont {Metelmann}\ and\ \citenamefont {Clerk}(2015)}]{metelmannNonreciprocalPhotonTransmission2015a}%
  \BibitemOpen
  \bibfield  {author} {\bibinfo {author} {\bibfnamefont {A.}~\bibnamefont {Metelmann}}\ and\ \bibinfo {author} {\bibfnamefont {A.~A.}\ \bibnamefont {Clerk}},\ }\href {https://doi.org/10.1103/PhysRevX.5.021025} {\bibfield  {journal} {\bibinfo  {journal} {Phys. Rev. X}\ }\textbf {\bibinfo {volume} {5}},\ \bibinfo {pages} {021025} (\bibinfo {year} {2015})}\BibitemShut {NoStop}%
\bibitem [{\citenamefont {Wineland}\ \emph {et~al.}(1994)\citenamefont {Wineland}, \citenamefont {Bollinger}, \citenamefont {Itano},\ and\ \citenamefont {Heinzen}}]{winelandSqueezedAtomicStates1994}%
  \BibitemOpen
  \bibfield  {author} {\bibinfo {author} {\bibfnamefont {D.~J.}\ \bibnamefont {Wineland}}, \bibinfo {author} {\bibfnamefont {J.~J.}\ \bibnamefont {Bollinger}}, \bibinfo {author} {\bibfnamefont {W.~M.}\ \bibnamefont {Itano}},\ and\ \bibinfo {author} {\bibfnamefont {D.~J.}\ \bibnamefont {Heinzen}},\ }\href {https://doi.org/10.1103/PhysRevA.50.67} {\bibfield  {journal} {\bibinfo  {journal} {Phys. Rev. A}\ }\textbf {\bibinfo {volume} {50}},\ \bibinfo {pages} {67} (\bibinfo {year} {1994})}\BibitemShut {NoStop}%
\bibitem [{\citenamefont {Pezz{\`e}}\ \emph {et~al.}(2018)\citenamefont {Pezz{\`e}}, \citenamefont {Smerzi}, \citenamefont {Oberthaler}, \citenamefont {Schmied},\ and\ \citenamefont {Treutlein}}]{pezzeQuantumMetrologyNonclassical2018}%
  \BibitemOpen
  \bibfield  {author} {\bibinfo {author} {\bibfnamefont {L.}~\bibnamefont {Pezz{\`e}}}, \bibinfo {author} {\bibfnamefont {A.}~\bibnamefont {Smerzi}}, \bibinfo {author} {\bibfnamefont {M.~K.}\ \bibnamefont {Oberthaler}}, \bibinfo {author} {\bibfnamefont {R.}~\bibnamefont {Schmied}},\ and\ \bibinfo {author} {\bibfnamefont {P.}~\bibnamefont {Treutlein}},\ }\href {https://doi.org/10.1103/RevModPhys.90.035005} {\bibfield  {journal} {\bibinfo  {journal} {Rev. Mod. Phys.}\ }\textbf {\bibinfo {volume} {90}},\ \bibinfo {pages} {035005} (\bibinfo {year} {2018})}\BibitemShut {NoStop}%
\end{thebibliography}%

\end{document}


\title{Supplemental Material for\\``Exact solutions of non-Markovian dissipative Ising models''}
\author{Mohammad Maghrebi}
\affiliation{Department of Physics and Astronomy, Michigan State University, East Lansing, Michigan 48824 USA}
\maketitle
In this Supplemental Material, we provide additional details on the results stated in the main text.
In \cref{sec:validity}, we present a simple proof of our stochastic representation of quantum dynamics.
We provide exact solutions for spin observables without local spin dissipation in \cref{sec:no_loss},
incorporate dephasing, relaxation, and pumping in \cref{sec:local_loss}, and treat coherence-generating dissipation in
\cref{sec:cg_sm}.
In \cref{sec:markov}, we derive an exact time-local equation for spin-only dynamics without local spin dissipation, perform the adiabatic elimination in the appropriate regime including spin dissipation, and further discuss dynamics under nonreciprocal interactions.
Finally, \cref{sec:squeezing} provides more details on our squeezing calculation in the main text.
\tableofcontents
\section{Stochastic approach}\label{sec:validity}

The stochastic representation used in the main text was first derived in  Ref.~\cite{kamarHybridQuantumClassicalStochastic2025a} using a combination of phase-space and functional integral methods. 
Here, we provide an alternative proof showing that this representation captures the exact evolution of arbitrary spin-boson observables.  For simplicity, we consider
one bosonic mode and real couplings,
\begin{equation}
    \hat H_{\rm int}
    =\frac{1}{\sqrt N}\sum_j g_j\sigma_j^z\hat x,
    \qquad
    \hat x=\frac{\hat a+\hat a^\dagger}{2},
    \qquad
    \hat y=\frac{\hat a-\hat a^\dagger}{2i},
\end{equation}
with $\hat x,\hat y$ the two quadratures of the cavity mode satisfying the commutation relation $[\hat x,\hat y]=i/2$. Let $\hat\rho(t)$ denote the physical density
matrix of the full system, and consider the expectation value of an observable,
\begin{equation}
    \langle \hat O \hat F\rangle_t
    \equiv
    \tr
    \left[\hat O\hat F\hat\rho(t)\right],
\end{equation}
where $\hat O$ and $\hat F$ are arbitrary spin and bosonic operators, respectively (we drop the tensor product for ease of notation). The Heisenberg equation gives
\begin{align}
    \partial_t \langle \hat O  \hat F\rangle_t
    &=\cdots + i\langle [\hat H_{\rm int}, \hat O \hat F]\rangle_t\nonumber \\
    &=\cdots +
    \frac{i}{2\sqrt N}\sum_j g_j
    \Big(
    \langle[\sigma_j^z,\hat O] \{\hat x,\hat F\}\rangle_t
    +\langle\{\sigma_j^z,\hat O\}[\hat x,\hat F]\rangle_t
    \Big),
    \label{eq:quantum_moment_hierarchy}
\end{align}
where the ellipses denote the contributions from the uncoupled spin and boson terms, and the identity $[\hat A\otimes\hat B,\hat C\otimes\hat D]=\frac12[\hat A,\hat C]\otimes\{\hat B,\hat D\}+\frac12\{\hat A,\hat C\}\otimes[\hat B,\hat D]$ is used in the second line. \Cref{eq:quantum_moment_hierarchy} then defines a hierarchy coupling
different spin+cavity operators.

To make contact with the stochastic description, we use the
Wigner--Weyl representation of the bosonic operators. A cavity
operator $\hat F$ is represented by its Weyl symbol
$f(x,y)\equiv(\hat F)_W$. For a fully symmetrized polynomial in the
quadratures, its Weyl symbol is obtained by replacing the operators
with classical variables.
Furthermore, the Weyl symbol of an operator product is given by the Moyal product of the corresponding Weyl symbols~\cite{polkovnikovPhaseSpaceRepresentation2010},
\begin{equation}
    (\hat A\hat B)_W
    =
    A_W
    \exp\left[
    \frac{i}{4}
    \left(
    \overleftarrow{\partial}_x\overrightarrow{\partial}_y
    -
    \overleftarrow{\partial}_y\overrightarrow{\partial}_x
    \right)
    \right]
    B_W.
\end{equation}
The factor $1/4$ follows from our quadrature normalization.
Specifically, we have
\begin{equation}
    (\hat x\hat F)_W
    =xf+\frac{i}{4}\partial_y f,
    \qquad
    (\hat F\hat x)_W
    =xf-\frac{i}{4}\partial_y f\,.
\end{equation}
Since $(\hat x)_W=x$ is linear, the expansion terminates after the first
derivative. The Weyl symbols of the (anti)commutators are then given by
\begin{equation}
    \left(\{\hat x,\hat F\}\right)_W=2 xf,
    \qquad
    \left([\hat x,\hat F]\right)_W
    =\frac{i}{2}\partial_y f.
    \label{eq:weyl_interaction_identities}
\end{equation}

We now independently define the stochastic averages
\begin{equation}\label{eq:Rt}
    \llangle \hat O \hat F\rrangle_t \equiv\E[f(x_t,y_t)\tr_S(\hat O \hat R_t)],
    \qquad
    \hat R_t\equiv \otimes_j\hat \rho_j(t),
\end{equation}
where the stochastic fields $x_t,y_t$ and single-spin operators $\hat \rho_j(t)$ evolve as (with the field $\psi(t)\equiv x_t+iy_t$)
\begin{align}
    i \dd \psi &= (\omega-i\kappa)\psi \dd t +\sqrt{\frac{\kappa}{2N}}\sum_j ( \dd W_j  + i\dd W_j')  \,, \\
    \dd \hat \rho_j &= {\cal L}_j(\hat \rho_j)\dd t  -\frac{ix_t}{\sqrt N} g_j[\sigma_j^z,\hat \rho_j(t)]\,\dd t
    +\frac{g_j}{2\sqrt{2\kappa}}
    \{\sigma_j^z,\hat \rho_j(t)\}\,\dd W_j.\label{eq:line2}
\end{align}
Here, $\mathcal L_j$ contains the local spin Hamiltonian and dissipation,
and $W_j,W_j'$ are independent standard Wiener processes.
Using It\^o's product rule
\begin{equation}\label{eq:Itoo}
    \dd(\hat R_t f)= (\dd\hat R_t)f+\hat R_t\dd f+\dd\hat R_t\,\dd f\,,
\end{equation}
together with the It\^o chain rule for $\dd f$, we have
\begin{equation}
\begin{aligned}
    \dd\llangle\hat O\hat F\rrangle_t
    ={}&\cdots+\frac{i}{\sqrt N}\sum_j g_j\,
    \E\!\left[\tr_S([\sigma_j^z,\hat O]\hat R_t)
    x_tf(x_t,y_t)\right]\dd t\\
    &-\sum_j\frac{g_j}{4\sqrt N}\,
    \E\!\left[\tr_S(\{\sigma_j^z,\hat O\}\hat R_t)
    \partial_y f(x_t,y_t)\right]\dd t.
\end{aligned}
\end{equation}
Again, the ellipses denote the contributions from the uncoupled terms. The first term on the rhs of this equation follows from the corresponding \textit{drift} term (proportional to $x_t$) in \cref{eq:line2}, while the second term on the rhs follows from the It\^o cross term $\dd\hat R_t\,\dd f$, since $\dd y_t|_{\rm noise}=-\sqrt{\kappa/(2N)}\sum_j\dd W_j$.
Notice that $\dd\hat\rho_j\,\dd\hat\rho_l=0$ for $j\ne l$, since $\dd W_j\dd W_l=0$, so no additional cross terms arise when differentiating the tensor product.
Using \cref{eq:weyl_interaction_identities} together with our definition in \cref{eq:Rt}, the above equation is then identical to \cref{eq:quantum_moment_hierarchy} upon the substitution $\langle\cdot\rangle\rightarrow\llangle\cdot\rrangle$. The uncoupled spin and bosonic terms contained in the ellipses also agree.
Therefore, the two sets of expectation values obey the same dynamical hierarchy. Assuming that the initial state is a product state of individual spins and the boson, the initial values also agree if we start from the same density matrix for each spin, and identify the initial state of the bosonic mode by its Wigner function in the stochastic approach. In short, the stochastic approach reproduces the expectation values of arbitrary spin-boson operators  with the bosonic observables substituted by their Weyl symbols.

Our construction also points to a (gauge) freedom in choosing the noise variables while preserving the same hierarchy. To show this in a more general setting, we first write the stochastic equations for generally complex $g_j$,
\begin{align}
\begin{split}
    i \dd \psi &= (\omega-i\kappa)\psi \dd t +\sqrt{\frac{\kappa}{2N}}\sum_j ( \dd W_j  + i\dd W_j')\,,   \\
    \dd \hat \rho_j &= {\cal L}_j(\hat \rho_j)\dd t  -\frac{i }{\sqrt N}\re (g_j\psi(t)) [\sigma_j^z,\hat \rho_j(t)]\,\dd t
    +\frac{1}{2\sqrt{2\kappa}}
    \{\sigma_j^z,\hat \rho_j(t)\}\,\re(g_j (\dd W_j+i \dd W_j'))\,.
\end{split}
\end{align}
A particularly useful transformation is $\dd W_j  + i\dd W_j' \to e^{i\theta_j}(\dd W_j  + i\dd W_j')$ in both lines of the above equations, for arbitrary fixed phases $\theta_j$. This rotation preserves the noise covariances and the It\^o cross terms. In particular, we can absorb the phase of $g_j$ into the noise variable by choosing $\theta_j = -\angle g_j$ to obtain
\begin{align}\label{eq:phase_abs}
\begin{split}
    i \dd \psi &= (\omega-i\kappa)\psi \dd t +\sqrt{\frac{\kappa}{2N}}\sum_j e^{-i \angle g_j}( \dd W_j  + i\dd W_j')\,,   \\
    \dd \hat \rho_j &= {\cal L}_j(\hat \rho_j)\dd t  -\frac{i }{\sqrt N} \re (g_j\psi(t))[\sigma_j^z,\hat \rho_j(t)]\,\dd t
    +\frac{|g_j|}{2\sqrt{2\kappa}}
    \{\sigma_j^z,\hat \rho_j(t)\}\, \dd W_j\,,
\end{split}
\end{align}
thus bringing the noise term in the spin dynamics to a simple form similar to \cref{eq:line2}.

\section{Exact solutions: No spin dissipation}\label{sec:no_loss}
Here, we explicitly compute the expectation values and correlation
functions of spin observables using our stochastic approach. In this section, we further assume no local spin dissipation, and set $\mathcal L_j=0$. For most calculations, we also consider a single bosonic mode for simplicity although generalization to multiple bosons is straightforward. As a concrete example, we assume that the spins are initially polarized along the positive $x$ direction, and the bosonic mode is uncorrelated with the spins, but otherwise in an arbitrary state described by its Wigner function.

We start by writing each spin's density matrix in the Bloch sphere representation,
\begin{equation}
 \hat\rho_j=\frac12(T_j\boldsymbol 1_j+X_j\sigma_j^x+Y_j\sigma_j^y+Z_j\sigma_j^z),
 \label{eq:sm_bloch}
\end{equation}
where the trace $T_j=\tr\hat\rho_j$ is generally not one, since the stochastic evolution is not trace preserving. With spins initially pointing along the $+x$ direction, we have $T_j(0)=X_j(0)=1$ and
$Y_j(0)=Z_j(0)=0$. For generally complex-valued $g_j$, the stochastic differential equations follow by plugging in \cref{eq:sm_bloch} in \cref{eq:phase_abs},
\begin{subequations}
\label{eq:sm_loss_bloch_0}
\begin{align}
    \dd X_l&=-h_l(s)Y_l\,\dd s,\\
    \dd Y_l&=h_l(s)X_l \,\dd s,\\
    \dd Z_l&= c_l T_l\,\dd W_l,\\
    \dd T_l&=c_lZ_l\,\dd W_l\,,
\end{align}
\end{subequations}
where $c_l=|g_l|/\sqrt{2\kappa}$ and
$h_l(t)=[g_l\psi(t)+g_l^*\psi(t)^*]/\sqrt N$.
These equations can then be solved exactly to obtain 
\begin{equation}
\begin{gathered}
 X_j(t)+iY_j(t)=e^{i\int_0^t\dd s\,h_j(s)},\\
 T_j(t)=e^{-c_j^2t/2}\cosh[c_jW_j(t)],\qquad
 Z_j(t)=e^{-c_j^2t/2}\sinh[c_jW_j(t)]\,.
\end{gathered}
 \label{eq:sm_trajectories}
\end{equation}
Also, the equation for $\psi$ can be solved explicitly to obtain
\begin{equation}
 \psi(t)=iG(t)\psi_0+\sqrt{\frac{\kappa}{2N}}
 \sum_l e^{-i\angle g_l}\int_0^t G(t-s)[\dd W_l(s)+i\dd W'_l(s)]\,.
 \label{eq:sm_cavity_solution}
\end{equation}
Here, $G(t)=-i\Theta(t)e^{-\varepsilon t}$ with $\varepsilon=\kappa+i\omega$ and $\Theta(t)$ the Heaviside function.
It will also be convenient to define its successive integrals
\begin{equation}
 \mathscr G(t)=\int_0^t\dd s\,G(s)
 =-i\frac{1-e^{-\varepsilon t}}{\varepsilon},\qquad
 \mathscr F(t)=\int_0^t\dd s\,\mathscr G(s)
 =-i\frac{\varepsilon t-1+e^{-\varepsilon t}}{\varepsilon^2}\,.
 \label{eq:cal_G}
\end{equation}

\subsection{Single-spin observables}
We first consider the expectation value of transverse components of spin $j$, which in the stochastic approach  reads
\begin{equation}\label{eq:YZ}
    \langle \sigma^x_j + i \sigma^y_j \rangle_t = \E [(X_j(t)+iY_j(t))\prod_{l\ne j} T_l(t)]\,.
\end{equation}
Again, we have included $T_l$ on all sites except $j$ since the corresponding density matrices are generally not trace-1.
We then proceed by writing the integrated field $h_j(s)$ in \cref{eq:sm_trajectories}
as a sum of the contribution from the initial bosonic field and those from the independent noise increments:
\begin{align}
 \int_0^t\dd s\,h_j(s)
 &=\frac1{\sqrt N}\left[i g_j\mathscr G(t)\psi_0+\mathrm{c.c.}\right]\nonumber\\
 &\quad+\sum_l\frac{\sqrt{2\kappa}}{N}\int_0^t
 \left\{\re[g_je^{-i\angle g_l}\mathscr G(t-s)]\dd W_l(s)
 -\im[g_je^{-i\angle g_l}\mathscr G(t-s)]\dd W'_l(s)\right\}.
 \label{eq:int_ht}
\end{align}
The average over the initial bosonic state gives precisely the factor defined
in the main text,
\begin{equation}
 \mathcal A_j(t)=\E_{\rm ini}\exp\!\left\{
 \frac{i}{\sqrt N}\left[i g_j\mathscr G(t)\psi_0+\mathrm{c.c.}\right]\right\}\,.
 \label{eq:sm_A}
\end{equation}
Here, we have defined $\E_{\rm ini}[\cdots] = \int \dd^2\psi_0 \mathscr W_0(\psi_0)\cdots$ where $\mathscr W_0(\psi_0)$ is the Wigner function of the bosonic mode in the initial state, and is not required to be either Gaussian or positive.
For a thermal state with mean occupation $\bar n$, we obtain $\mathcal A_j(t)=\exp[-(\bar n+1/2)|g_j\mathscr G(t)|^2/N]$; we set $\bar n=0$ for the vacuum state.

To compute the noise averages, we use
\begin{equation}
 \E\exp\!\left[\int_0^t f(s)\dd W(s)\right]
 =\exp\!\left[\frac12\int_0^t\dd s\,f(s)^2\right]\,,
 \label{eq:martingale}
\end{equation}
for deterministic, possibly complex $f$. Writing the hyperbolic functions
in \cref{eq:sm_trajectories} as sums of exponentials, we can bring various stochastic averages into this form.

Denoting by $h_l^{(j)}$ the contribution to $h_j$ from the noise increments with the index $l$, we have
\begin{equation}
\begin{aligned}
 \E_l\!\left[e^{i\int_0^t\dd s\,h_l^{(j)}(s)}\right]
 &=e^{-\Lambda_{jj}(t)/N},\\
 \E_l\!\left[T_l(t)e^{i\int_0^t\dd s\,h_l^{(j)}(s)}\right]
 &=e^{-\Lambda_{jj}(t)/N}\cos\Theta_{jl}(t),\\
 \E_l\!\left[Z_l(t)e^{i\int_0^t\dd s\,h_l^{(j)}(s)}\right]
 &=i e^{-\Lambda_{jj}(t)/N}\sin\Theta_{jl}(t)\,,
\end{aligned}
 \label{eq:sm_single_noise_average}
\end{equation}
where, as in the main text, we define $\Theta_{jl}, \Lambda_{jl}$ (allowing different indices on the latter):
\begin{equation}
 \Theta_{jl}(t)=\frac1N\re[g_jg_l^*\mathscr F(t)],\qquad
 \Lambda_{jl}(t)=\frac{\kappa}{N}\re(g_jg_l^*)
 \int_0^t\dd s\,|\mathscr G(s)|^2.
 \label{eq:sm_theta_lambda}
\end{equation}
In particular, $\Theta_{jl}$ need not equal $\Theta_{lj}$ for complex couplings. Using the above equations, we recover the expression in the main text,
\begin{equation}
 \langle\sigma_j^x+i\sigma_j^y\rangle_t
 =\mathcal A_j(t)e^{-\Lambda_{jj}(t)}
 \prod_{l\ne j}\cos\Theta_{jl}(t).
 \label{eq:sm_no_loss_coherence}
\end{equation}
The longitudinal expectation values are trivial to calculate in general, and vanish for the initial state considered here.

\subsection{Two-spin correlations}

For two spins, the product of the transverse components contains the phase
$\int_0^t[h_j(s)\pm h_k(s)]\dd s$. Thus the same calculation applies with
$g_j$ replaced by $g_j\pm g_k$, and with the trace factors multiplied for
$l\notin\{j,k\}$. The initial bosonic state now contributes the factor
\begin{equation}
 \mathcal A_{jk}^{\pm}(t)\equiv\E_{\rm ini}
 \exp\!\left\{\frac{i}{\sqrt N}
 [i(g_j\pm g_k)\mathscr G(t)\psi_0+\mathrm{c.c.}]\right\}.
 \label{eq:sm_Apair}
\end{equation}
Putting all the pieces together, we obtain
\begin{equation}
 \langle(\sigma_j^x+i\sigma_j^y)(\sigma_k^x\pm i\sigma_k^y)\rangle_t
 =\mathcal A_{jk}^{\pm}(t)
 e^{-\Lambda_{jj}(t)-\Lambda_{kk}(t)\mp2\Lambda_{jk}(t)}
 \prod_{l\notin\{j,k\}}\cos[\Theta_{jl}(t)\pm\Theta_{kl}(t)].
 \label{eq:sm_no_loss_pair}
\end{equation}
For the correlation function between transverse and longitudinal components on different sites, the calculation proceeds in a similar way, with the trace on site $k$ replaced with $Z_k$. We then obtain
\begin{equation}
 \langle(\sigma_j^x+i\sigma_j^y)\sigma_k^z\rangle_t
 =i\mathcal A_j(t)e^{-\Lambda_{jj}(t)}\sin\Theta_{jk}(t)
 \prod_{l\notin\{j,k\}}\cos\Theta_{jl}(t).
 \label{eq:sigma_pm_2}
\end{equation}

\subsection{Bosonic observables}

Bosonic observables and their equal-time correlations with spins can be obtained from our stochastic approach too; cf. \cref{sec:validity}. The expectation value of the bosonic operator can also be obtained directly from the Heisenberg--Langevin equation:
\begin{equation}
 \langle\hat a(t)\rangle
 =e^{-(\kappa+i\omega)t}\langle\hat a(0)\rangle
 -\frac{i}{2\sqrt N}\sum_l g_l^*
 \int_0^t\dd s\,e^{-(\kappa+i\omega)(t-s)}
 \langle\sigma_l^z(s)\rangle.
 \label{eq:sm_mean_cavity}
\end{equation}
For mixed spin--boson correlations, the Heisenberg--Langevin equation generally
involves unequal-time spin correlations as well. Alternatively, our stochastic approach can
compute these mixed observables directly.
These correlators are not needed for the results presented in the main text, and will not be reported here.

\section{Exact solutions: Local spin dissipation in \texorpdfstring{$z$}{z} basis}\label{sec:local_loss}

In this section, we include local spin dissipation in the $z$ basis. We derive closed equations for the stochastic averages entering the  spin observables. We  interpret these averages in terms of spin-flip histories and obtain \mainequation{14} in the main text. We also derive the two-spin correlators which will be used for calculating spin squeezing.

Considering dephasing, spontaneous emission, and pumping, with rates
$\Gamma_\phi,\Gamma_\downarrow,\Gamma_\uparrow$, respectively, the local dissipator takes the form
\begin{equation}
 \mathcal L_j=\Gamma_\phi\mathcal D[\sigma_j^z]
 +\Gamma_\downarrow\mathcal D[\sigma_j^-]
 +\Gamma_\uparrow\mathcal D[\sigma_j^+],\qquad
 \mathcal D[\hat L](\hat \rho)=\hat L\hat \rho \hat L^\dagger-\tfrac12\{\hat L^\dagger \hat L,\hat \rho\}.
\end{equation}
It is convenient to define
\begin{equation}
    \Gamma=\frac{\Gamma_\downarrow+\Gamma_\uparrow}{2},
    \qquad
    \eta=\frac{\Gamma_\downarrow-\Gamma_\uparrow}{2},
    \qquad
    \Gamma'=2\Gamma_\phi+\Gamma.
\end{equation}
With the local spin dissipation and allowing for generally complex couplings, the stochastic equations now take the form
\begin{subequations}
\label{eq:sm_loss_bloch}
\begin{align}
    \dd X_l&=[-h_l(s)Y_l-\Gamma'X_l]\,\dd s,\\
    \dd Y_l&=[h_l(s)X_l-\Gamma'Y_l]\,\dd s,\\
    \dd Z_l&=(-2\eta T_l-2\Gamma Z_l)\,\dd s+c_l T_l\,\dd W_l,\\
    \dd T_l&=c_lZ_l\,\dd W_l,
\end{align}
\end{subequations}
where again
\(
    c_l=\frac{|g_l|}{\sqrt{2\kappa}},
\)
and
\(
    h_l(s)=\frac{g_l\psi(s)+g_l^*\psi(s)^*}{\sqrt N}.
\)
The pairs $(X_l,Y_l)$ and $(Z_l,T_l)$ remain decoupled.
The stochastic solution for the transverse components is given by
\begin{equation}
    X_j(t)+iY_j(t)
    =
    e^{-\Gamma't+i\int_0^t \dd s\,h_j(s)}
    [X_j(0)+iY_j(0)],
\end{equation}
which differs from the solution without spin dissipation only by an overall exponential decay. However, the stochastic solutions for $Z_l,T_l$ are not generally available in closed form. We instead derive closed equations for their stochastic moments.
As in \cref{sec:no_loss}, we initially take the spins to be polarized along $+x$, with the bosonic mode in an arbitrary, uncorrelated initial
state.

\subsection{Single-spin observable from stochastic moments}

For a given  spin $j$ and a fixed time $t$, the contribution
of the noise with index $l$ to the integrated longitudinal field is
\begin{equation}
    \int_0^t \dd s\,h_l^{(j)}(s)
    =
    \int_0^t
    \left[
        a_l(s)\,\dd W_l(s)
        +a_l'(s)\,\dd W_l'(s)
    \right],
\end{equation}
where
\begin{equation}
\begin{aligned}
    a_l(s)
    &=
    \frac{\sqrt{2\kappa}}{N}
    \operatorname{Re}\!\left[
        g_j e^{-i\angle g_l}\mathscr G(t-s)
    \right],\\
    a_l'(s)
    &=
    -\frac{\sqrt{2\kappa}}{N}
    \operatorname{Im}\!\left[
        g_j e^{-i\angle g_l}\mathscr G(t-s)
    \right].
\end{aligned}
\end{equation}
Here, the dependence of $a_l,a_l'$ (and $M_l$ defined below) on $j$ and the
fixed final time $t$ is left implicit.

Since $Z_l,T_l$ depend only on $W_l$, the average over $W_l'$
can be performed separately. To compute the remaining average, we introduce
the \textit{martingale}
\begin{equation}
    M_l(s)
    =
    \exp\!\left[
        i\int_0^s a_l(u)\,\dd W_l(u)
        +\frac12\int_0^s \dd u\,a_l(u)^2
    \right].
\end{equation}
For fixed $t$, this satisfies the stochastic differential equation
\begin{equation}
    \dd M_l(s)=i\,a_l(s)M_l(s)\,\dd W_l(s),
    \qquad M_l(0)=1.
\end{equation}
Using this definition, we can write the stochastic average that appears in our calculation as
\begin{equation}
\begin{aligned}
    \mathbb E_l\!\left[
        T_l(t)e^{i\int_0^t \dd s\,h_l^{(j)}(s)}
    \right]
    &=
    e^{-\frac12\int_0^t \dd s\,
    [a_l(s)^2+a_l'(s)^2]}
    \mathbb E_l[T_l(t)M_l(t)] = e^{-\Lambda_{jj}(t)/N}
    \mathbb E_l[T_l(t)M_l(t)].
\end{aligned}
\end{equation}
To find a closed set of equations, we define the stochastic moments
\begin{equation}
    p_l(s;t)=\mathbb E_l[T_l(s)M_l(s)],
    \qquad
    q_l(s;t)=\mathbb E_l[Z_l(s)M_l(s)].
    \label{eq:sm_loss_moments}
\end{equation}
Applying It\^o's rule to the two products and averaging, we find
\begin{equation}
\label{eq:sm_loss_pq}
    \partial_s
    \begin{pmatrix}
        p_l(s;t)\\
        q_l(s;t)
    \end{pmatrix}
    =
    \begin{pmatrix}
        0 & ik_{jl}(t-s)\\
        ik_{jl}(t-s)-2\eta & -2\Gamma
    \end{pmatrix}
    \begin{pmatrix}
        p_l(s;t)\\
        q_l(s;t)
    \end{pmatrix}\,,
\end{equation}
with $p_l(0;t)=1$ and $q_l(0;t)=0$. Here, we have used
\begin{equation}
 c_la_l(s)=k_{jl}(t-s), \qquad k_{jl}(u)=\dot\Theta_{jl}(u)
 =\frac1N\re[g_jg_l^*\mathscr G(u)]\,.
 \label{eq:sm_phase_rate}
\end{equation}
The above equations provide a linear system of ordinary differential equations that can be used to solve for $p_l(s;t)$.
Finally, putting different pieces together, we obtain the spin coherence as
\begin{equation}
    \left\langle\sigma_j^x+i\sigma_j^y\right\rangle_t
    =
    \mathcal A_j(t)e^{-\Gamma't-\Lambda_{jj}(t)}
    \prod_{l\ne j}p_{jl}(t),
    \qquad
    p_{jl}(t)\equiv p_l(t;t).
    \label{eq:sm_loss_coherence}
\end{equation}

As a final remark, we note that our analysis extends to multiple bosonic modes by replacing the kernel with the sum of the contributions from each mode,
\begin{equation}
 k_{jl}(u)=\frac1N\sum_\alpha
 \re[g_{\alpha j}g_{\alpha l}^*\mathscr G_\alpha(u)].
\end{equation}

\subsection{Phase accumulation}
\Cref{eq:sm_loss_pq} admits an interpretation in terms of the phase accumulation. To see this, we first define
\begin{equation}
    P_{jl}^{\uparrow/\downarrow}(s;t)
    =\frac{p_l(s;t)\pm q_l(s;t)}{2}
    =
    \frac12\mathbb E_l\!\left[
        (T_l(s)\pm Z_l(s))M_l(s)
    \right].
\end{equation}
Equation~\eqref{eq:sm_loss_pq} then becomes
\begin{equation}
    \frac{\partial}{\partial s}
    \begin{pmatrix}
        P_{jl}^{\uparrow}\\
        P_{jl}^{\downarrow}
    \end{pmatrix}
    =
    \begin{pmatrix}
        ik_{jl}(t-s)-\Gamma_\downarrow
        & \Gamma_\uparrow\\
        \Gamma_\downarrow
        & -ik_{jl}(t-s)-\Gamma_\uparrow
    \end{pmatrix}
    \begin{pmatrix}
        P_{jl}^{\uparrow}\\
        P_{jl}^{\downarrow}
    \end{pmatrix},
    \label{eq:sm_loss_forward}
\end{equation}
with $P_{jl}^{\uparrow/\downarrow}(0;t)=1/2$.

Next, consider a classical two-state process $z_l(s)=\pm1$,
with downward and upward transition rates
$\Gamma_\downarrow$ and $\Gamma_\uparrow$, respectively,
and equally likely initial orientations.
It is easy to see that the quantities
\begin{equation}
 \E_{\rm flips}\!\left[
 e^{i\int_0^s\dd u\,k_{jl}(t-u)z_l(u)}
 ;\,z_l(s)=\pm1\right]=
 \E_{\rm flips}\!\left[
 e^{i\int_0^s\dd u\,k_{jl}(t-u)z_l(u)}
 \,\middle|\,z_l(s)=\pm1\right]\Pr[z_l(s)=\pm1]
 \label{eq:sm_loss_forward_average}
\end{equation}
obey the same equations and initial conditions as
$P_{jl}^{\uparrow/\downarrow}(s;t)$, respectively: The off-diagonal terms transfer weight between the two
orientations, while the imaginary diagonal terms account for
the phase accumulated in the evolution.

At $s=t$, summing over the final orientation therefore gives
\begin{equation}
    p_{jl}(t)
    =
    \mathbb E_{\rm flips}\!\left[e^{i\phi_{jl}(t)}\right],
    \qquad
    \phi_{jl}(t)
    =
    \int_0^t \dd s\,k_{jl}(t-s)z_l(s).
    \label{eq:sm_loss_phase}
\end{equation}
Here $\phi_{jl}(t)$ is the phase accumulated by spin $j$ due to
a given spin-flip history of spin $l$. The phase interpretation thus follows directly from the stochastic equations.

The kernel in \cref{eq:sm_loss_forward} depends on both $s$ and the final time $t$. To obtain \mainequation{14} in the main text with a single time variable, we instead condition the phase average on the initial orientation and define
\begin{equation}
    p_{jl}^{\uparrow/\downarrow}(t)
    =
    \mathbb E_{\rm flips}\!\left[
        e^{i\phi_{jl}(t)}
        \,\middle|\,z_l(0)=\pm1
    \right].
    \label{eq:sm_loss_conditional_definition}
\end{equation}
Since no phase has accumulated initially,
$p_{jl}^{\uparrow/\downarrow}(0)=1$, and, for the initial state considered,
\(
    p_{jl}(t)
    =
    (p_{jl}^{\uparrow}(t)+p_{jl}^{\downarrow}(t))/2.
\)

To find the evolution of $p_{jl}^{\uparrow/\downarrow}$, we consider $p_{jl}^{\uparrow/\downarrow}(t+\delta t)$ and split the accumulated phase between an initial interval $[0,\delta t]$ and the remaining interval of length $t$. We note that
\begin{equation}
\begin{aligned}
    \phi_{jl}(t+\delta t)
    &=\int_0^{t+\delta t}\dd s\,k_{jl}(t+\delta t-s)z_l(s)\\
    &=\int_0^{\delta t}\dd s\,k_{jl}(t+\delta t-s)z_l(s)
    +\int_0^t\dd s\,k_{jl}(t-s)z_l(s+\delta t),
\end{aligned}
\end{equation}
where we have written the second term in the last equation by shifting the integration variable.
Since the flips form a Markov process with constant rates, the phase factor from the second term, conditioned on \(z_l(\delta t)=\pm1\), averages to \(p_{jl}^{\uparrow/\downarrow}(t)\), respectively. We can also replace the first term by \(k_{jl}(t)z_l(0)\delta t\), with an error of \(O(\delta t^2)\) after averaging: a flip during this interval changes the phase by $O(\delta t)$ and occurs with probability $O(\delta t)$.
For an initially up spin, we therefore obtain
\begin{equation}
 p^\uparrow_{jl}(t+\delta t)
 =e^{ik_{jl}(t)\delta t}
 \left[(1-\Gamma_\downarrow\delta t)p^\uparrow_{jl}(t)
 +\Gamma_\downarrow\delta t\,p^\downarrow_{jl}(t)\right]+O(\delta t^2).
\end{equation}
Writing the corresponding equation for $p^\downarrow_{jl}(t+\delta t)$ and expanding to first order in $\delta t$, we finally obtain
\begin{equation}
    \frac{\dd}{\dd t}
    \begin{pmatrix}
        p_{jl}^{\uparrow}\\
        p_{jl}^{\downarrow}
    \end{pmatrix}
    =
    \begin{pmatrix}
        ik_{jl}(t)-\Gamma_\downarrow
        & \Gamma_\downarrow\\
        \Gamma_\uparrow
        & -ik_{jl}(t)-\Gamma_\uparrow
    \end{pmatrix}
    \begin{pmatrix}
        p_{jl}^{\uparrow}\\
        p_{jl}^{\downarrow}
    \end{pmatrix}\,,
    \label{eq:sm_loss_conditional_matrix}
\end{equation}
and recover \mainequation{14} of the main text.

For a general initial product state, the conditional equations
and their unit initial values remain unchanged. The initial spin
populations instead enter through
\begin{equation}
    p_{jl}(t)
    =
    \frac{1+\langle\sigma_l^z\rangle_0}{2}
    p_{jl}^{\uparrow}(t)
    +
    \frac{1-\langle\sigma_l^z\rangle_0}{2}
    p_{jl}^{\downarrow}(t).
\end{equation}

In the absence of pumping, \(\Gamma_\uparrow=0\), the matrix in \cref{eq:sm_loss_conditional_matrix} becomes upper triangular, and the equations can be integrated explicitly. In this case, we find
\begin{equation*}
p_{jl}^{\downarrow}(t)=e^{-i\Theta_{jl}(t)}.
\end{equation*}
Substituting this into the equation for \(p_{jl}^{\uparrow}\), we also obtain
\begin{equation*}
p_{jl}^{\uparrow}(t)
=
e^{-\Gamma_\downarrow t+i\Theta_{jl}(t)}
\left[
1+\Gamma_\downarrow\int_0^t \dd s\,
e^{\Gamma_\downarrow s-2i\Theta_{jl}(s)}
\right].
\end{equation*}
For spins initially along $+x$, the average of these functions gives
\begin{equation}
\begin{aligned}
    p_{jl}(t)
    =\frac{p_{jl}^{\uparrow}(t)+p_{jl}^{\downarrow}(t)}{2}=
    \frac{
        e^{-i\Theta_{jl}(t)}
        +e^{-\Gamma_\downarrow t+i\Theta_{jl}(t)}
    }{2}
    +
    \frac{\Gamma_\downarrow}{2}
    \int_0^t \dd\tau\,
    e^{-\Gamma_\downarrow\tau
    +i\Theta_{jl}(t)-2i\Theta_{jl}(t-\tau)}.
    \label{eq:sm_loss_no_pumping}
\end{aligned}
\end{equation}
This expression has a simple interpretation in terms of phase accumulation.
Without pumping, each spin can decay at most once. The three terms on the rhs of the above equation describe
an initially down spin, an initially up spin that remains up, and one that
decays at time $\tau$, integrated over all possible decay times. The latter
history accumulates the phase
\begin{equation}
 \int_0^\tau\dd s\,k_{jl}(t-s)
 -\int_\tau^t\dd s\,k_{jl}(t-s)
 =\Theta_{jl}(t)-2\Theta_{jl}(t-\tau)\,,
\end{equation}
consistent with the above equation.
When spontaneous emission
is also absent, $p_{jl}(t)=\cos\Theta_{jl}(t)$, and the coherence reduces to
\cref{eq:sm_no_loss_coherence} multiplied by $e^{-2\Gamma_\phi t}$.

\subsection{Two-spin correlations}

Similar stochastic calculations give the two-spin correlators. Correlations between transverse components of spins $j,k$
involve the phase $\int_0^t[h_j(s)\pm h_k(s)]\dd s$.
Since the noise from spin $l$ contributes to both fields, the corresponding
phases add or subtract. The calculation thus follows that of the
single-spin observable closely, yielding
\begin{equation}
 \langle(\sigma_j^x+i\sigma_j^y)(\sigma_k^x\pm i\sigma_k^y)\rangle_t
 =\mathcal A_{jk}^{\pm}(t)
 e^{-2\Gamma't-\Lambda_{jj}(t)-\Lambda_{kk}(t)\mp2\Lambda_{jk}(t)}
 \prod_{l\notin\{j,k\}}p_{jk;l}^{\pm}(t).
 \label{eq:sm_loss_pair}
\end{equation}
Here, $p_{jk;l}^{\pm}(t)$ is the final value $p_l(t;t)$ obtained by
solving \cref{eq:sm_loss_pq} with the replacement
\begin{equation}
 k_{jl}(t-s)\longrightarrow
 k_{jl}(t-s)\pm k_{kl}(t-s),
 \label{eq:sm_pair_kernel}
\end{equation}
together with the same initial conditions.

For a transverse-longitudinal correlator between spins $j,k$ respectively, the factor $T_k(t)$ on site $k$ is replaced by $Z_k(t)$, while the phase factor due to the spin $j$ remains unchanged. The corresponding average is therefore the second component of the vector in \cref{eq:sm_loss_pq},
\begin{equation}
 q_{jk}(t)\equiv q_k(t;t)
 =\E_k[Z_k(t)M_k(t)],
 \label{eq:sm_endpoint_q}
\end{equation}
where $M_k$ is defined as before (with the index $j$ implicit).
We then obtain
\begin{equation}
 \langle(\sigma_j^x+i\sigma_j^y)\sigma_k^z\rangle_t
 =\mathcal A_j(t)e^{-\Gamma't-\Lambda_{jj}(t)}q_{jk}(t)
 \prod_{l\notin\{j,k\}}p_{jl}(t).
 \label{eq:sm_loss_mixed}
\end{equation}

To calculate the new factors that appear in the above equations, it is convenient to use the picture based on the phase accumulation, and condition it on the initial spin orientation.
For the transverse correlators, define
\begin{equation}
 p_{jk;l}^{\pm,\uparrow/\downarrow}(t)
 =
 \E_{\rm flips}\!\left[
 e^{i[\phi_{jl}(t)\pm\phi_{kl}(t)]}
 \,\middle|\,z_l(0)=\pm1
 \right].
 \label{eq:sm_pair_conditional}
\end{equation}
Both phases are evaluated along the same history of spin $l$. (The sign between the phases is independent of the sign specifying the initial orientation.)
These quantities obey \cref{eq:sm_loss_conditional_matrix} with
$k_{jl}(t)\to k_{jl}(t)\pm k_{kl}(t)$ and initial values
$p_{jk;l}^{\pm,\uparrow/\downarrow}(0)=1$, and $p_{jk;l}^{\pm}(t)
 =(p_{jk;l}^{\pm,\uparrow}(t)
       +p_{jk;l}^{\pm,\downarrow}(t))/{2}$.

For the mixed correlator, the insertion of $Z_k(t)$ gives an additional factor for the final spin orientation in the phase average:
\begin{equation}
 q_{jk}(t)
 =\E_{\rm flips}\!\left[z_k(t)e^{i\phi_{jk}(t)}\right].
\end{equation}
Paralleling the above analysis, we can define
\begin{equation}
 q_{jk}^{\uparrow/\downarrow}(t)
 =\E_{\rm flips}\!\left[z_k(t)e^{i\phi_{jk}(t)}
 \,\middle|\,z_k(0)=\pm1\right].
 \label{eq:sm_endpoint_conditional}
\end{equation}
A similar analysis shows that these quantities obey the same matrix equation as
$p_{jk}^{\uparrow/\downarrow}$, but with
$q_{jk}^{\uparrow}(0)=1$ and $q_{jk}^{\downarrow}(0)=-1$.
For spins initially along $+x$, the two initial orientations have equal
weights, \(
 q_{jk}(t)
 =({q_{jk}^{\uparrow}(t)+q_{jk}^{\downarrow}(t)})/{2}.
\)

Finally, the longitudinal observables obey independent local rate equations, yielding
\begin{equation}
 \langle\sigma_j^z(t)\rangle
 =-\frac{\eta}{\Gamma}(1-e^{-2\Gamma t}),
 \qquad
 \langle\sigma_j^z(t)\sigma_k^z(t)\rangle
 =\langle\sigma_j^z(t)\rangle\langle\sigma_k^z(t)\rangle
 \quad(j\ne k).
 \label{eq:sm_longitudinal}
\end{equation}

\section{Exact solutions: Coherence-generating dissipation}
\label{sec:cg_sm}

In this section, we further consider local dissipation that generates coherence while retaining exact solvability. We first construct these processes through measurement in the $z$ basis followed by preparation of a prescribed state. We then provide the exact stochastic solution.

\subsection{Measurement in \texorpdfstring{$z$}{z} basis and preparation}
Suppressing the site index, consider a
projective measurement of $\sigma^z$, followed by a unitary $\hat U_\uparrow$
or $\hat U_\downarrow$ depending on the outcome. Averaging over the outcomes
gives a measurement-and-preparation channel~\cite{horodeckiGeneralEntanglementBreaking2003},
\begin{equation}
 \mathcal E(\hat \rho)=\sum_{s=\uparrow,\downarrow}\hat U_s\hat P_s\hat \rho \hat P_s \hat U_s^\dagger
 =\rho_{\uparrow\uparrow}|\chi_\uparrow\rangle\langle\chi_\uparrow|
+\rho_{\downarrow\downarrow}|\chi_\downarrow\rangle\langle\chi_\downarrow|,
 \label{eq:cg_sm_channel}
\end{equation}
where
\begin{equation}
    \hat P_s=\lvert s\rangle\langle s\rvert,
    \qquad
    \lvert\chi_s\rangle=\hat U_s\lvert s\rangle.
\end{equation}
We can allow any pair of normalized pure output states.
For spin $j$ in the full system, the same local operation gives
\begin{equation}
 \mathcal E_j(\hat\rho)
 =\sum_{s=\uparrow,\downarrow}
 |\chi_s\rangle_j\langle\chi_s|\otimes{}_j\langle s|\hat\rho|s\rangle_j,
 \label{eq:cg_sm_many_spin_channel}
\end{equation}
where ${}_j\langle s|\hat\rho|s\rangle_j$ is an unnormalized density matrix involving other spins and bosons. The tensor product with spin $j$ is understood  in the original order.

If these operations occur according to a Poisson process with rate \(r\), then averaging over this process gives, over an interval \(\mathrm dt\), 
\begin{equation}
    \hat\rho(t+\dd t)
    =(1-r\,\dd t)\hat\rho(t)
    +r\,\dd t\,\mathcal E[\hat\rho(t)]
    +O(\dd t^2)\,,
    \label{eq:cg_sm_poisson}
\end{equation}
where we have suppressed the site index again.
The local master equation then becomes
\begin{equation}
    \dot{\hat\rho}\big|_{\rm loc}
    =r\bigl[\mathcal E(\hat\rho)-\hat\rho\bigr]
    =\mathcal D[\hat L_\uparrow]\hat\rho
    +\mathcal D[\hat L_\downarrow]\hat\rho,
    \label{eq:cg_sm_poisson_generator}
\end{equation}
with
\begin{equation}
    \hat L_s=\sqrt r\,\lvert\chi_s\rangle\langle s\rvert,
    \qquad
    \mathcal D[\hat L]\hat\rho
    =\hat L\hat\rho \hat L^\dagger-\frac12\{\hat L^\dagger \hat L,\hat\rho\}.
\end{equation}
Applying these operations independently on each site gives the generator
$r\sum_j[\mathcal E_j(\hat\rho)-\hat\rho]$. Complete positivity and trace preservation follow directly from this construction.

We can also allow different rates for the two jump operators,
\begin{equation}
    \hat L_\uparrow
    =\sqrt{r_\uparrow}\,\lvert\chi_\uparrow\rangle\langle\uparrow\rvert,
    \qquad
    \hat L_\downarrow
    =\sqrt{r_\downarrow}\,\lvert\chi_\downarrow\rangle\langle\downarrow\rvert.
    \label{eq:cg_sm_general_jumps}
\end{equation}
Equal rates describe the projective measurement protocol above. Unequal rates give a broader family of local jump processes,
including ordinary spontaneous emission.
Writing the output states as
\begin{equation}
    \lvert\chi_s\rangle
    =\alpha_s\lvert\uparrow\rangle+\beta_s\lvert\downarrow\rangle,
    \qquad
    |\alpha_s|^2+|\beta_s|^2=1\,,
    \label{eq:cg_sm_output_states}
\end{equation}
the local generator becomes
\begin{equation}
    \mathcal L_{\rm loc}(\hat\rho)
    =\sum_{s=\uparrow,\downarrow}
    r_s\left[
    \rho_{ss}\lvert\chi_s\rangle\langle\chi_s\rvert
    -\frac12\{\hat P_s,\hat\rho\}
    \right].
    \label{eq:cg_sm_feedback_generator}
\end{equation}

The above generator determines the dynamics of the populations in the $z$ basis immediately:
\begin{align}
    \label{eq:cg_sm_populations}
        \dot\rho_{\uparrow\uparrow}
        =-r_\uparrow|\beta_\uparrow|^2\rho_{\uparrow\uparrow}
        +r_\downarrow|\alpha_\downarrow|^2\rho_{\downarrow\downarrow},\qquad
        \dot\rho_{\downarrow\downarrow}
        =-\dot\rho_{\uparrow\uparrow}.
    \end{align}
These equations contain no coherences, underlying the exact solvability.  Defining the effective population-transfer rates
\begin{equation}
    \Gamma_\downarrow=r_\uparrow|\beta_\uparrow|^2,
    \qquad
    \Gamma_\uparrow=r_\downarrow|\alpha_\downarrow|^2,
    \label{eq:cg_sm_population_rates}
\end{equation}
together with
\begin{equation}
    \Gamma=\frac{\Gamma_\downarrow+\Gamma_\uparrow}{2},
    \qquad
    \eta=\frac{\Gamma_\downarrow-\Gamma_\uparrow}{2},
\end{equation}
the $Z, T$ variables evolve locally as
\begin{equation}
    \dot T\big|_{\rm loc}=0,
    \qquad
    \dot Z\big|_{\rm loc}=-2\eta T-2\Gamma Z.
    \label{eq:cg_sm_TZ_local}
\end{equation}
Again, we have defined $T=\tr\hat\rho$ and
$Z=\tr(\sigma^z\hat\rho)$.

Similarly, since $C=X+iY=2\rho_{\downarrow\uparrow}$, the off-diagonal
element of \cref{eq:cg_sm_feedback_generator} gives
\begin{equation}
 \dot C\big|_{\rm loc}=-\Gamma'C+aT+bZ,
 \qquad
 \Gamma'=\frac{r_\uparrow+r_\downarrow}{2},\qquad
 a,b=r_\uparrow\beta_\uparrow\alpha_\uparrow^*
 \mathbin{\pm}r_\downarrow\beta_\downarrow\alpha_\downarrow^*\,.
 \label{eq:cg_sm_coefficients}
\end{equation}
Notice that the equation for $C$ does not depend on $C^*$, making it exactly solvable with the $T,Z$ terms acting as a source. Additional dephasing, spontaneous emission, and pumping are
incorporated by adding their contributions to these coefficients.

For equal rates and
$\lvert\chi_\uparrow\rangle=\lvert\chi_\downarrow\rangle=\lvert+x\rangle$,
these expressions give
\begin{equation}
    \Gamma=\frac r2,\qquad
    \eta=0,\qquad
    \Gamma'=r,\qquad
    a=r,\qquad b=0.
    \label{eq:cg_sm_reset_example}
\end{equation}
Choosing instead $\lvert\chi_\uparrow\rangle=\lvert+x\rangle$ and
$\lvert\chi_\downarrow\rangle=\lvert-x\rangle$ gives the same decay rates
but $a=0,\ b=r$: the generated coherence then depends on the population
imbalance.

\subsection{Spin observables from stochastic moments}
\label{sec:cg_sm_solution}

We take the local rates and preparation states to be the same at each site.
We allow an arbitrary initial product state of the spins, uncorrelated with the initial bosonic state.
In our stochastic approach, the resulting equations are given by  (with $C_j=X_j+iY_j$)
\begin{subequations}
    \label{eq:cg_sm_stochastic}
    \begin{align}
        \dd T_j
        &=c_jZ_j\,\dd W_j,\\
        \dd Z_j
        &=(-2\eta T_j-2\Gamma Z_j)\,\dd t
        +c_jT_j\,\dd W_j,\\
        \dd C_j
        &=[ih_j(t)-\Gamma']C_j\,\dd t
        +(aT_j+bZ_j)\,\dd t.
    \end{align}
\end{subequations}
We use the phase-rotated noises in \cref{eq:phase_abs}, so these equations
also apply to complex couplings, with $c_j=|g_j|/\sqrt{2\kappa}$ and
$h_j(t)=[g_j\psi(t)+g_j^*\psi(t)^*]/\sqrt N$.

Since $T_j,Z_j$ evolve autonomously, the coherence equation can be
integrated with these variables as a source:
\begin{align}
    C_j(t)
    =
    e^{-\Gamma't+i\int_0^t \dd s\,h_j(s)}C_j(0)
    +\int_0^t \dd\tau\,
    e^{-\Gamma'(t-\tau)+i\int_\tau^t \dd s\,h_j(s)}
    \bigl[aT_j(\tau)+bZ_j(\tau)\bigr].
    \label{eq:cg_sm_C_solution}
\end{align}
The single integral accounts for the continuous source of coherence, and does not
restrict the number of physical reset events.
To compute $\langle\sigma_j^x+i\sigma_j^y\rangle_t$, we multiply
\cref{eq:cg_sm_C_solution} by $\prod_{l\ne j}T_l(t)$ and average, as in
\cref{eq:YZ}. The source term therefore requires the average
\begin{equation}
 \E\!\left[
 e^{i\int_\tau^t\dd u\,h_j(u)}
 [aT_j(\tau)+bZ_j(\tau)]\prod_{l\ne j}T_l(t)
 \right].
 \label{eq:cg_sm_source_average}
\end{equation}
The phase in this equation accumulates only over $[\tau,t]$, but the cavity field still depends on noise at earlier times. Thus resetting a spin does not erase the memory of the bosonic mode.
We can write this phase as
\begin{align}
 \int_\tau^t\dd u\,h_j(u)
 &=\frac1{\sqrt N}[i g_j {\mathscr K}(t,\tau)\psi_0+\mathrm{c.c.}]+\frac{\sqrt{2\kappa}}N\sum_l\int_0^t
 \re\!\left[g_je^{-i\angle g_l} {\mathscr K}(t-s,\tau-s)
 (\dd W_l(s)+i\dd W'_l(s))\right]\,,
 \label{eq:cg_sm_phase}
\end{align}
where we have defined
\begin{equation}
    \mathscr K(u,v)=\mathscr G(u)-\mathscr G(v)\,.
    \label{eq:cg_sm_interval_kernel}
\end{equation}
Note that $\mathscr G(v)=0$ for $v<0$.
The kernel $\mathscr K(t-s,\tau-s)$ generally remains nonzero for $s<\tau$.

To compute the average in \cref{eq:cg_sm_source_average}, we proceed as before. The coefficient of $\dd W_l(s)$ in \cref{eq:cg_sm_phase} is
\begin{equation}
 f_l(s)=\frac{\sqrt{2\kappa}}{N}
 \re[g_je^{-i\angle g_l}\mathscr K(t-s,\tau-s)].
 \label{eq:cg_sm_noise_amplitude}
\end{equation}
We introduce the exponential martingale
\begin{equation}
    M_l(s)=
    \exp\left[
    i\int_0^s f_l(u)\,\dd W_l(u)
    +\frac12\int_0^s \dd u\,f_l(u)^2
    \right].
    \label{eq:cg_sm_martingale}
\end{equation}
This satisfies $\dd M_l(s)=if_l(s)M_l(s)\,\dd W_l(s)$.
The dependence of $f_l$ and $M_l$ on the fixed $j,t,\tau$ is suppressed.
As in \cref{eq:sm_loss_moments,eq:sm_loss_pq}, separating the Gaussian
factor and averaging over $W_l'$ leaves the moments
\begin{equation}
    p_l(s;t,\tau)=\mathbb E[T_l(s)M_l(s)],
    \qquad
    q_l(s;t,\tau)=\mathbb E[Z_l(s)M_l(s)].
    \label{eq:cg_sm_moments}
\end{equation}
The kernel governing the phase accumulation is now given by
\begin{equation}
 k_{jl}(s;t,\tau)=\frac1N\re[g_jg_l^*\mathscr K(t-s,\tau-s)]
 =k_{jl}(t-s)-k_{jl}(\tau-s),
 \label{eq:cg_sm_phase_rate}
\end{equation}
where $k_{jl}(t)$ is defined in
\cref{eq:sm_phase_rate}.
Applying It\^o's rule to both products gives
\begin{equation}
    \frac{\dd}{\dd s}
    \begin{pmatrix}p_l(s;t,\tau)\\q_l(s;t,\tau)\end{pmatrix}
    =
    \begin{pmatrix}
        0&ik_{jl}(s;t,\tau)\\
        ik_{jl}(s;t,\tau)-2\eta&-2\Gamma
    \end{pmatrix}
    \begin{pmatrix}p_l(s;t,\tau)\\q_l(s;t,\tau)\end{pmatrix}.
    \label{eq:cg_sm_moment_equations}
\end{equation}
For normalized product initial states,
\begin{equation}
    p_l(0;t,\tau)=1,\qquad q_l(0;t,\tau)=Z_l(0).
\end{equation}
These equations hold for $0\le s\le t$, and fixed $j,t,\tau$.

We combine the contribution from the initial bosonic state and the Gaussian dephasing factor into
\begin{equation}
 \mathcal B_j(t,\tau)
 =\E_{\rm ini}\exp\!\left\{\frac{i}{\sqrt N}
 [i g_j\mathscr K(t,\tau)\psi_0+\mathrm{c.c.}]\right\}
 \exp\!\left[-\frac{\kappa|g_j|^2}{N}
 \int_0^t\dd s\,|\mathscr K(t-s,\tau-s)|^2\right].
 \label{eq:cg_sm_bosonic_factor}
\end{equation}
Specifically, $\mathcal B_j(t,0)=\mathcal A_j(t)e^{-\Lambda_{jj}(t)}$.

On site $j$, the source contains $T_j(\tau),Z_j(\tau)$, so the moments are evaluated at $s=\tau$. The Wiener increments after $\tau$ are independent of the source, and averaging over them gives $M_j(\tau)$ by the martingale property. Thus

\begin{equation}
 \E\!\left[T_j(\tau)M_j(t)\right]
 =\E\!\left[T_j(\tau)M_j(\tau)\right],\qquad
 \E\!\left[Z_j(\tau)M_j(t)\right]
 =\E\!\left[Z_j(\tau)M_j(\tau)\right].
\end{equation}
The other sites retain their trace factors at time $t$.
Using the moments defined in \cref{eq:cg_sm_moments}, the single-spin coherence is
\begin{align}
    \langle\sigma_j^x+i\sigma_j^y\rangle_t
    &=
    C_j(0)e^{-\Gamma't}\mathcal B_j(t,0)
    \prod_{l\ne j}p_l(t;t,0)
    \nonumber\\
    &\quad+
    \int_0^t \dd\tau\,
    e^{-\Gamma'(t-\tau)}\mathcal B_j(t,\tau)
    \left[
    ap_j(\tau;t,\tau)+bq_j(\tau;t,\tau)
    \right]
    \prod_{l\ne j}p_l(t;t,\tau).
    \label{eq:cg_sm_spin_coherence}
\end{align}
Thus the source contributes one ordinary time integral, with two-dimensional equations for the remaining averages at each fixed $t,\tau$. Exact solvability follows because the populations evolve independently of the coherences.
The calculation also extends to correlations, with each transverse spin operator introducing at most one integral over the time at which its coherence is generated. The remaining averages obey two-dimensional equations, with non-Markovian memory entering through $\mathscr K$.

\section{Markovian limit}\label{sec:markov}
In this section, we first derive an exact convolutionless, or time-local, equation for the reduced spin density matrix in the absence of local spin dissipation. In the Markovian limit, we recover an Ising model with correlated dephasing and, for complex couplings, nonreciprocal interactions, while allowing for local spin dissipation. Finally, we obtain spin observables and show how nonreciprocity enters the dynamics of an arbitrary subsystem.

\subsection{Spin-only dynamics}

We first integrate out the bosonic mode exactly in the absence of a transverse field and local spin dissipation. For a single mode, the interaction Hamiltonian is given by
\begin{equation}
 \hat H_{\rm int}=\frac12(\hat L^\dagger \hat a+\hat L \hat a^\dagger),
 \qquad  \hat L\equiv\frac1{\sqrt N}\sum_j g_j^*\sigma_j^z.
\end{equation}
Following the phase-space representation in Eqs.~(16)--(19) of
Ref.~\cite{kamarHybridQuantumClassicalStochastic2025a}, the evolution of the reduced density matrix of the spins can be written as
\begin{equation}
 \hat\rho_S(t)=\left\langle\mathcal T_t
 \exp\!\left[\int_0^t\dd u\,
 \big(\mathcal L_S+ \mathcal L_{\rm int}(u)\big)\right]
 \right\rangle_B\hat\rho_S(0),
 \label{eq:sm_gaussian_propagator}
\end{equation}
where $T_t$ denotes time ordering, and $\mathcal L_S$ describes the spin-only Hamiltonian and local spin dissipation. The interaction superoperator is given by
\begin{equation}
 \mathcal L_{\rm int}(t)
 =i[\psi(t)\mathcal S^*+\psi(t)^*\mathcal S
       +\phi(t)\mathcal T^*+\phi(t)^*\mathcal T],
 \label{eq:sm_gaussian_Lint}
\end{equation}
with
\begin{equation}
\begin{aligned}
 \mathcal S(\hat\rho)&=-\tfrac12[\hat L,\hat \rho],&
 \mathcal S^*(\hat\rho)&=-\tfrac12[\hat L^\dagger,\hat \rho],\\
 \mathcal T(\hat\rho)&=-\tfrac12\{\hat L,\hat \rho\},&
 \mathcal T^*(\hat\rho)&=-\tfrac12\{\hat L^\dagger,\hat \rho\}.
\end{aligned}
 \label{eq:sm_gaussian_ST}
\end{equation}
The star on a superoperator indicates the conjugated couplings in $\hat L$.
The average $\langle\cdot\rangle_B$ is the phase-space integral over the classical fields $\psi(t),\phi(t)$. For a bosonic mode initially in the vacuum and uncorrelated with the spins, the average is over a generally complex Gaussian distribution with zero mean and the correlations
\begin{equation}
\begin{aligned}
 \langle\psi(u)\psi(v)^*\rangle_B
 &
 =\frac{i}{2}[G(u-v)-G(v-u)^*] ,\\
 \langle\psi(u)\phi(v)^*\rangle_B&=\tfrac i2G(u-v),&
 \langle\phi(u)\psi(v)^*\rangle_B&=\tfrac i2G(v-u)^*\,,
\end{aligned}
 \label{eq:sm_gaussian_contractions}
\end{equation}
with $G(t)=-i\Theta(t)e^{-i\omega t-\kappa t}$.
All anomalous correlations and $\langle\phi(u)\phi(v)^*\rangle_B$
vanish.

Now setting $\mathcal L_S=0$, and noting that $[\hat L,\hat L^\dagger]=0$, we find that the superoperators in \cref{eq:sm_gaussian_ST} all commute, and the time ordering
in \cref{eq:sm_gaussian_propagator} can be dropped. Defining
$\Psi(t)=\int_0^t\dd u\,\psi(u)$ and
$\Phi(t)=\int_0^t\dd u\,\phi(u)$, we
then have
\begin{equation}
\begin{aligned}
 \hat\rho_S(t)=\langle e^{i\mathcal A(t)}\rangle_B\,\hat\rho_S(0)
 =e^{-\frac12\langle\mathcal A(t)^2\rangle_B}\,\hat\rho_S(0),
 \qquad \mathcal A(t)=\Psi(t)\mathcal S^*+\Psi(t)^*\mathcal S
           +\Phi(t)\mathcal T^*+\Phi(t)^*\mathcal T.
\end{aligned}
 \label{eq:sm_gaussian_average}
\end{equation}
Using \cref{eq:sm_gaussian_contractions}, we have
\begin{equation}
 \langle\Psi\Psi^*\rangle_B=-\im\mathscr F(t),\qquad
 \langle\Psi\Phi^*\rangle_B=\tfrac i2\mathscr F(t),\qquad
 \langle\Psi^*\Phi\rangle_B=\tfrac i2\mathscr F(t)^*\,,
 \label{eq:sm_integrated_contractions}
\end{equation}
where $\mathscr F(t)$ is defined in \cref{eq:cal_G}.
In the last two expressions, the factor $i$ is not conjugated: the
retarded and advanced contractions both carry the prefactor $i/2$.
We thus obtain
\begin{equation}
 \hat\rho_S(t)=e^{\mathcal K(t)}\hat\rho_S(0), \qquad \mathcal K(t)
 =\im\mathscr F(t)\,\mathcal S^*\mathcal S
 -\tfrac i2\mathscr F(t)\,\mathcal S^*\mathcal T
 -\tfrac i2\mathscr F(t)^*\,\mathcal S\mathcal T^* .
 \label{eq:sm_gaussian_cumulant}
\end{equation}
Finally, substituting for superoperators from \cref{eq:sm_gaussian_ST}  yields an evolution operator in a more familiar form,
\begin{equation}
 \mathcal K(t)(\hat\rho)
 =-\frac{i}{4}\re\mathscr F(t)[\hat L^\dagger \hat L,\hat \rho]
 -\frac12\im\mathscr F(t)\mathcal D[\hat L](\hat \rho),
 \qquad
 \mathcal D[\hat L](\hat \rho)=\hat L\hat \rho \hat L^\dagger-\frac12\{\hat L^\dagger \hat L,\hat \rho\}.
 \label{eq:sm_integrated_generator}
\end{equation}
Since all (super)operators commute, $\mathcal K(t)$ commutes with its time derivative too, and we find the time-local equation
\begin{equation}
 \dot{\hat\rho}_S=-\frac{i}{4}\re\mathscr G(t)[\hat L^\dagger \hat L,\hat\rho_S]
 -\frac12\im\mathscr G(t)\mathcal D[\hat L](\hat\rho_S),
 \label{eq:sm_time_local}
\end{equation}
comprising the Hamiltonian as well as dissipators with time-dependent coefficients. Specifically, the coefficient of the dissipator can become negative, a signature of non-Markovianity \cite{devegaDynamicsNonMarkovianOpen2017}.
The above equation is exact for $\mathcal L_S=0$. Local longitudinal
fields and dephasing in the $z$ basis can also be included because their
superoperators commute with those above. Relaxation, pumping, and resetting do not generally commute with the interaction superoperators, so adding them directly to \cref{eq:sm_time_local} does not give the exact non-Markovian dynamics.

At large cavity loss and/or frequency, $\mathscr G(t)$ quickly approaches a constant after a short transient,
\(
 \mathscr G(t)\simeq-{i}/({\kappa+i\omega})
\). This gives the effective Hamiltonian and dissipator within the adiabatic elimination
\begin{equation}
\hat H_S=- \frac{1}{4}\frac{\omega}{\omega^2+\kappa^2} \hat L^\dagger \hat L, \qquad {\cal D}_{S} = \frac{1}{2}\frac{\kappa}{\omega^2+\kappa^2}{\cal D}[\hat L].
\end{equation}

For multiple modes, the contributions from each mode add, giving the effective spin model reported in \mainequations{16}{17} of the main text:
\begin{align}
 \hat H_S&=\sum_{j<l}J_{jl}\sigma_j^z\sigma_l^z,\qquad
 \mathcal D_S^{\rm corr}(\hat\rho_S)=\sum_{j,l}\Gamma_{jl}
 \left(\sigma_j^z\hat\rho_S\sigma_l^z-\frac12\{\sigma_j^z\sigma_l^z,\hat\rho_S\}\right),
 \label{eq:sm_ising_model}\\
 J_{jl}&=-\frac1{2N}\sum_\alpha
 \frac{\omega_\alpha}{\omega_\alpha^2+\kappa_\alpha^2}
 \re(g_{\alpha j}^*g_{\alpha l}),\qquad
 \Gamma_{jl}=\frac1{2N}\sum_\alpha
 \frac{\kappa_\alpha}{\omega_\alpha^2+\kappa_\alpha^2}g_{\alpha j}^*g_{\alpha l}.
 \label{eq:sm_J_Gamma}
\end{align}
The factors of $1/N$ are included in the definition of $J_{jl}$ and $\Gamma_{jl}$.
The matrix $\Gamma_{jl}$ is Hermitian and positive semidefinite, ensuring a
physical Lindblad dissipator, while its imaginary part leads to nonreciprocity.

At long times, the coefficients in the time-local equation, \cref{eq:sm_time_local}, always approach the same values upon adiabatic elimination; however, the exact and the adiabatically eliminated dynamics need not agree. The spins can accumulate a finite phase or undergo dephasing during the initial transient, which change the initial conditions for the subsequent evolution. Unless the transient effect is small or eventually washed out by relaxation, the two approaches give different results.

\subsection{Spin observables with local dissipation}
To calculate spin observables in the adiabatically eliminated model including relaxation and pumping, we return to the exact stochastic averages derived in \cref{sec:local_loss} and take their adiabatic limit. The phase kernels then become time independent,
\begin{equation}
    k_{jl}(t)\longrightarrow 2J_{jl}^{\rm eff},
    \qquad
    J_{jl}^{\rm eff}=J_{jl}-\im\Gamma_{jl},
    \label{eq:sm_Jeff}
\end{equation}
and the resulting matrix in
\cref{eq:sm_loss_conditional_matrix} now has constant coefficients.
For spins initially polarized along $+x$, solving this equation and
averaging over the two initial orientations gives
\begin{equation}
    p_{jl}(t)=e^{-\Gamma t}
    \left[
        \cosh(S_{jl}t)
        +\frac{\Gamma}{S_{jl}}\sinh(S_{jl}t)
    \right],
    \qquad
    S_{jl}=
    \sqrt{\Gamma^2-4J_{jl}^{\rm eff}
    (J_{jl}^{\rm eff}+i\eta)} \equiv S(J^{\rm eff}_{jl}).
    \label{eq:p_markovian}
\end{equation}
In the same limit,
$\Lambda_{jk}(t)\to2\re\Gamma_{jk}\,t$, and $\mathcal A_j(t)\approx 1$ for bosons initially in their vacuum state. Plugging in these
expressions into \cref{eq:sm_loss_coherence}, we obtain
\begin{equation}
    \langle\sigma_j^x+i\sigma_j^y\rangle_t
    =
    e^{-(\Gamma'+2\Gamma_{jj})t}
    \prod_{l\ne j}p_{jl}(t).
    \label{eq:sigma_plus_spontaneous}
\end{equation}
This and the following equations recover, but also extend the results of
Refs.~\cite{foss-feigNonequilibriumDynamicsArbitraryrange2013a,mcdonaldExactSolutionsInteracting2022} to include correlated dephasing through $\Gamma_{jl}$ and nonreciprocal couplings through $J_{jl}^{\rm eff}$.

The two-spin correlations follow from
\cref{eq:sm_loss_pair,eq:sm_loss_mixed}. For transverse components,
the phase due to an external spin $l$ involves the sum or difference
of its couplings to spins $j$ and $k$, as in
\cref{eq:sm_pair_kernel}. The factors
$p_{jk;l}^{\pm}$ are then obtained from \cref{eq:p_markovian} by replacing
$J_{jl}^{\rm eff}$ with
$J_{jl}^{\rm eff}\pm J_{kl}^{\rm eff}$. Explicitly,
\begin{equation}
    p_{jk;l}^{\pm}(t)=e^{-\Gamma t}
    \left[
        \cosh(S_{jk;l}^{\pm}t)
        +\frac{\Gamma}{S_{jk;l}^{\pm}}
        \sinh(S_{jk;l}^{\pm}t)
    \right], \qquad S_{jk;l}^{\pm} =S(J^{\rm eff}_{jl}\pm J^{\rm eff}_{kl}).
    \label{eq:sm_markov_pair_factor}
\end{equation}
The transverse correlators are therefore
\begin{equation}
\begin{aligned}
    &\langle(\sigma_j^x+i\sigma_j^y)
    (\sigma_k^x\pm i\sigma_k^y)\rangle_t=
    e^{-\left[
        2\Gamma'
        +2(\Gamma_{jj}+\Gamma_{kk}\pm2\re\Gamma_{jk})
    \right]t}
    \prod_{l\notin\{j,k\}}p_{jk;l}^{\pm}(t).
\end{aligned}
    \label{eq:sm_markov_pair_loss}
\end{equation}

For the mixed transverse-longitudinal correlator,
we use the same matrix equation
\eqref{eq:sm_loss_conditional_matrix}, but with initial values
$(1,-1)$ instead of $(1,1)$ as described around \cref{eq:sm_endpoint_conditional}. Solving with these initial conditions
gives
\begin{equation}
    q_{jk}(t)=
    \frac{2iJ_{jk}^{\rm eff}-2\eta}{S_{jk}}
    e^{-\Gamma t}\sinh(S_{jk}t).
    \label{eq:sm_markov_q}
\end{equation}
Substituting this result into \cref{eq:sm_loss_mixed}, we find
\begin{equation}
    \langle(\sigma_j^x+i\sigma_j^y)\sigma_k^z\rangle_t
    =
    e^{-(\Gamma'+2\Gamma_{jj})t}
    q_{jk}(t)
    \prod_{l\notin\{j,k\}}p_{jl}(t).
    \label{eq:sm_markov_mixed_loss}
\end{equation}

In the absence of local spin dissipation,
$\Gamma=\eta=\Gamma'=0$, and the hyperbolic functions reduce to
trigonometric functions. The above expressions then give
\begin{align}
    \langle(\sigma_j^x+i\sigma_j^y)
    (\sigma_k^x\pm i\sigma_k^y)\rangle_t
    &=
    e^{-2(\Gamma_{jj}+\Gamma_{kk}\pm2\re\Gamma_{jk})t}
    \prod_{l\notin\{j,k\}}
    \cos[2(J_{jl}^{\rm eff}\pm J_{kl}^{\rm eff})t],
    \label{eq:sm_markov_pair}\\
    \langle\sigma_j^y\sigma_k^z\rangle_t
    &=
    e^{-2\Gamma_{jj}t}
    \sin(2J_{jk}^{\rm eff}t)
    \prod_{l\notin\{j,k\}}
    \cos(2J_{jl}^{\rm eff}t).
    \label{eq:sm_markov_yz}
\end{align}
The dynamics of the nonreciprocal Ising model with no local spin dissipation takes a rather simple form \cite{seifCompressedSensingMeasurement2021a}.
The nonreciprocity between spins $j,k$ becomes manifest by inspecting the transverse-longitudinal correlators: exchanging the transverse and longitudinal operators in \cref{eq:sm_markov_yz}, their correlators now depend on $J_{kj}^{\rm eff}$ instead of $J_{jk}^{\rm eff}$.

\subsection{Subsystem dynamics with nonreciprocity}\label{sec:sybsystem}

In this subsection, we consider the nonreciprocal Ising model and show that
couplings across the boundary of a subsystem enter its dynamics only through
the effective coupling
\begin{equation}
    J_{jl}^{\rm eff}=J_{jl}-\operatorname{Im}\Gamma_{jl},
    \qquad j\in \mathcal C,\quad l\notin \mathcal C,
    \label{eq:Jeff_boundary}
\end{equation}
where $\mathcal C$ denotes the subsystem. The dissipative
couplings with both indices in $\mathcal C$ can, of course, appear independently.

To see this, consider an arbitrary operator $\hat O_{\mathcal C}$ supported on $\mathcal C$. Terms
acting entirely within its complement do not contribute to its Heisenberg equation of
motion, while terms acting entirely within $\mathcal C$ define the intrinsic
dynamics via the adjoint Liouvillian $\mathcal L_{\mathcal C}^\ddagger$. It therefore remains to consider a pair
$j\in \mathcal C$, $l\notin \mathcal C$. The corresponding contribution is
\begin{align}
    \mathcal L_{jl}^\ddagger(\hat O_{\mathcal C})
    ={}&
    iJ_{jl}[\sigma_j^z,\hat O_{\mathcal C}]\sigma_l^z
    \nonumber\\
    &+
    \left[
    \Gamma_{jl}
    \left(
    \hat O_{\mathcal C}\sigma_j^z-\frac{1}{2}\{\hat O_{\mathcal C},\sigma_j^z\}
    \right)
    +
    \Gamma_{lj}
    \left(
    \sigma_j^z\hat O_{\mathcal C}-\frac{1}{2}\{\hat O_{\mathcal C},\sigma_j^z\}
    \right)
    \right]\sigma_l^z .
    \label{eq:boundary_adjoint}
\end{align}
Using the Hermiticity of the dissipation matrix,
$\Gamma_{lj}=\Gamma_{jl}^*$, the second line simplifies to
\begin{equation}
    -i\operatorname{Im}\Gamma_{jl}
    [\sigma_j^z,\hat O_{\mathcal C}]\sigma_l^z.
\end{equation}
We thus find
\begin{equation}
    \frac{\dd\hat O_{\mathcal C}}{\dd t}
    =
    \mathcal L_{\mathcal C}^\ddagger(\hat O_{\mathcal C})
    +
    i\sum_{\substack{j\in \mathcal C\\ l\notin \mathcal C}}
    J_{jl}^{\rm eff}
    [\sigma_j^z,\hat O_{\mathcal C}]\sigma_l^z .
    \label{eq:subsystem_heisenberg}
\end{equation}
Therefore, $\operatorname{Re}\Gamma_{jl}$ cancels for couplings across the
boundary, while $J_{jl}$ and $\operatorname{Im}\Gamma_{jl}$ enter only through
$J_{jl}^{\rm eff}$.

Next, we note that \cref{eq:subsystem_heisenberg} generates operators of the form
\begin{equation}
    \hat O_{\mathcal C}\prod_{l\in X}\sigma_l^z,
    \qquad X\subseteq \overline{\mathcal C}.
\end{equation}
Applying the Liouvillian again preserves this form, changing the operator within $\mathcal C$ and adding or removing factors of $\sigma_l^z$ in its complement. The resulting hierarchy is therefore closed, and
its dependence on the couplings across the boundary remains entirely through
$J_{jl}^{\rm eff}$.
The same closure property holds in the presence of local decay and pumping in
the $z$ basis, and for the measurement-and-preparation processes of
\cref{sec:cg_sm}. Indeed, their adjoint dissipators map the space spanned by
the identity and $\sigma_l^z$ into itself. Their local population dynamics in the complement must therefore be retained,
but the cross-boundary dynamics only enters through the effective coupling in \cref{eq:Jeff_boundary}.

This property indicates that the subsystem dynamics is indistinguishable from one with the
same internal dynamics in $\mathcal C$, reciprocal Ising couplings of strength
$J_{jl}^{\rm eff}$ across its boundary, and no cross-boundary dissipation.
Within the complement, Ising interactions and dephasing act trivially on this
hierarchy and can be omitted, while relaxation, pumping and their coherence generating variants should be kept. For the same initial spin state, the two models therefore give identical reduced dynamics in $\mathcal C$, while their full states can differ. The combination of coherent and dissipative couplings in the Heisenberg equation parallels the analysis of Ref.~\cite{metelmannNonreciprocalPhotonTransmission2015a}. Here, the closed hierarchy ensures that the full subsystem dynamics depends on  cross-boundary couplings only through $J_{jl}^{\rm eff}$ for $j\in \mathcal C, l\notin \mathcal C$, without requiring the reverse couplings to vanish.  

\section{Spin squeezing}\label{sec:squeezing}

In this section, we derive the collective moments and numerical equations used for the squeezing example in the main text. We consider $N$ spins coupled to one bosonic mode with a uniform real coupling strength $g$. The boson is initially in its vacuum state, and the spins are polarized along $+x$.
The local pumping and relaxation rates are equal,
$\Gamma_\uparrow=\Gamma_\downarrow=\gamma$, so
$\eta=0$, $\Gamma=\gamma$, and $\Gamma'=\gamma+2\Gamma_\phi$.
The expectation value of the spin operator remains along $x$, and
$\langle\sigma_j^z\sigma_k^z\rangle=0$ for distinct sites.

Combining the contribution from the initial state and dephasing due to cavity loss into $D(t)$, we have ($\varepsilon=\kappa+i\omega$)
\begin{align}
 D(t)=-\log\mathcal A_j(t)+\Lambda_{jj}(t)
 =\frac{g^2}{N}\re\!\left[
 \frac{\varepsilon t-1+e^{-\varepsilon t}}{\varepsilon^2}\right].
 \label{eq:sm_squeeze_D}
\end{align}
The remaining factors are determined from the matrix equation
\eqref{eq:sm_loss_conditional_matrix}, with the same kernel $k(t)$,
\begin{align}
 k(t)=\frac{g^2}{N}\re\mathscr G(t)
 =-\frac{g^2}{N(\omega^2+\kappa^2)}
 \left[\omega-e^{-\kappa t}(\omega\cos\omega t+\kappa\sin\omega t)\right],
 \label{eq:sm_squeeze_k}
\end{align}
and $\Gamma_\uparrow=\Gamma_\downarrow=\gamma$.
Let us first consider $p_{jl}^{\uparrow/\downarrow}$ with the initial values $(1,1)$.
Since $k(t)$ is real and the two flip rates are equal, the two solutions are
complex conjugates. Suppressing the site labels, we write
\begin{equation}
 p^\uparrow=p+id,\qquad p^\downarrow=p-id.
 \label{eq:sm_squeeze_p_definition}
\end{equation}
The average over the initial orientation gives $p_{jl}=p$.
Adding and subtracting the two conditional equations gives
$\dot p=-kd$ and $\dot d=kp-2\gamma d$, with $p(0)=1,d(0)=0$.

To compute $\langle\sigma_j^y\sigma_k^z\rangle$, we also need the phase
average weighted by the final orientation $z_k(t)$, as in
\cref{eq:sm_endpoint_conditional}. The corresponding conditional averages
$q_{jk}^{\uparrow/\downarrow}$ obey the same matrix equation, with the initial values $(1,-1)$. Their solutions satisfy
$q^\downarrow=-(q^\uparrow)^*$, so we write
\begin{equation}
 q^\uparrow=v-iu,\qquad q^\downarrow=-v-iu,\qquad
 q_{jk}=\frac{q^\uparrow+q^\downarrow}{2}=-iu.
 \label{eq:sm_squeeze_q_definition}
\end{equation}
This gives $\dot u=-kv$, $\dot v=ku-2\gamma v$, with
$u(0)=0,v(0)=1$.

Finally, the correlators of transverse spin components use the sum or difference of the phase kernels in \cref{eq:sm_pair_kernel,eq:sm_loss_pair}.
For uniform coupling, these are $2k(t)$ and zero, respectively.
In the first case, we define $p_2,d_2$ as in
\cref{eq:sm_squeeze_p_definition}, but solve the equations with $k$
replaced by $2k$. In the second case, no phase is accumulated, and the
factor is unity. Together, the six real equations used in the calculation are
\begin{equation}
 \begin{aligned}
 \dot p&=-k(t)d,&\qquad\dot d&=k(t)p-2\gamma d,\\
 \dot u&=-k(t)v,&\dot v&=k(t)u-2\gamma v,\\
 \dot p_2&=-2k(t)d_2,&\dot d_2&=2k(t)p_2-2\gamma d_2,
 \end{aligned}
 \qquad
 (p,d,u,v,p_2,d_2)_{t=0}=(1,0,0,1,1,0).
 \label{eq:sm_squeeze_odes}
\end{equation}
The single-spin coherence follows from \cref{eq:sm_loss_coherence},
\begin{equation}
 m_x(t)\equiv\langle\sigma_j^x\rangle
 =e^{-\Gamma't-D(t)}p(t)^{N-1}.
 \label{eq:sm_squeeze_mx}
\end{equation}
For the two-spin correlations, \cref{eq:sm_loss_pair} gives
\begin{align}
\begin{split}
 \langle(\sigma_j^x+i\sigma_j^y)(\sigma_k^x+i\sigma_k^y)\rangle
 &=e^{-2\Gamma't-4D}p_2^{N-2},\\
 \langle(\sigma_j^x+i\sigma_j^y)(\sigma_k^x-i\sigma_k^y)\rangle
 &=e^{-2\Gamma't}.
\end{split}
\end{align}
These expressions then give
\begin{equation}
 c_{yy}(t)\equiv\langle\sigma_j^y\sigma_k^y\rangle
 =\frac12e^{-2\Gamma't}
 \left[1-e^{-4D(t)}p_2(t)^{N-2}\right].
 \label{eq:sm_squeeze_cyy}
\end{equation}
Similarly, taking the imaginary part of \cref{eq:sm_loss_mixed}, with
$q_{jk}=-iu$, gives
\begin{equation}
 c_{yz}(t)\equiv\langle\sigma_j^y\sigma_k^z\rangle
 =-e^{-\Gamma't-D(t)}u(t)p(t)^{N-2},
 \qquad c_{zz}(t)=0,
 \label{eq:sm_squeeze_correlators}
\end{equation}
for $j\ne k$.
In the absence of spin flips, these equations give
$p=\cos\Theta$, $p_2=\cos(2\Theta)$, and $u=-\sin\Theta$ with $\Theta(t)=\int_0^t\dd s\,k(s)$, in agreement with \cref{eq:sm_no_loss_pair,eq:sigma_pm_2}. For constant $k$, the solutions also follow explicitly from
\cref{eq:p_markovian,eq:sm_markov_q}, with $2J_{jl}^{\rm eff}=k$
and $\eta=0$. With both spin flips and non-Markovian memory,
we integrate \cref{eq:sm_squeeze_odes} numerically.

With $\hat S_\mu=\frac12\sum_j\sigma_j^\mu$, the relevant collective moments are
\begin{equation}
 \langle \hat S_x\rangle=\frac N2m_x,\quad
 V_{yy}=\frac N4[1+(N-1)c_{yy}],\quad
 V_{zz}=\frac N4,\quad
 V_{yz}=\frac12\langle\{\hat S_y,\hat S_z\}\rangle
 =\frac{N(N-1)}4c_{yz}\,,
\end{equation}
while $\langle \hat S_y\rangle=\langle \hat S_z\rangle=0$. Here, $V$ is the covariance
matrix in the $y$--$z$ plane. The Wineland parameter is then given by
\begin{equation}
 \xi_R^2=\frac{N V_{\min}}{\langle \hat S_x\rangle^2},\qquad
 V_{\min}=\frac{V_{yy}+V_{zz}
 -\sqrt{(V_{yy}-V_{zz})^2+4V_{yz}^2}}2.
 \label{eq:sm_wineland}
\end{equation}
Values below unity indicate metrologically useful spin squeezing~\cite{winelandSqueezedAtomicStates1994,pezzeQuantumMetrologyNonclassical2018}.

We also consider the adiabatically eliminated spin model with
\begin{equation}
 \hat H_S=-\frac{\chi}{N}\hat S_z^2,\qquad
 \mathcal D_S^{\rm corr}=\frac{2\kappa g^2}{N(\omega^2+\kappa^2)}\mathcal D[\hat S_z],
 \qquad\chi=\frac{g^2\omega}{\omega^2+\kappa^2}.
 \label{eq:sm_squeeze_eliminated}
\end{equation}
An irrelevant constant has been dropped from $\hat H_S$. Equivalently, the eliminated results follow from
\cref{eq:sm_squeeze_odes} and the correlators in
\crefrange{eq:sm_squeeze_mx}{eq:sm_squeeze_correlators} with
\begin{equation}
 k(t)\longrightarrow-\frac{\chi}{N},\qquad
 D(t)\longrightarrow\frac{\kappa g^2}{N(\omega^2+\kappa^2)}t.
\end{equation}
In particular, there will be no collective dephasing in the absence of cavity loss.

\bibliographystyle{apsrev4-2}
\bibliography{exact_2}